\documentclass[aps,prc,amsfonts,preprintnumbers,superscriptaddress,showpacs,nofootinbib]{revtex4}

\usepackage{graphics}
\usepackage{amssymb}
\usepackage{psfrag}
\usepackage{epsfig}
\usepackage{epsf}
\usepackage{float}

\newcommand{\fet}[1]{\mbox{\boldmath $#1$}}

\newcommand{\beq}{\begin{equation}}
\newcommand{\eeq}{\end{equation}}
\newcommand{\beqa}{\begin{eqnarray}}
\newcommand{\eeqa}{\end{eqnarray}}
\newcommand{\nn}{\nonumber \\ }

\begin{document}

\title{Chiral three-nucleon force at N$^4$LO I: 
Longest-range contributions}

\author{H.~Krebs}
\email[]{Email: hermann.krebs@rub.de}
\affiliation{Institut f\"ur Theoretische Physik II, Ruhr-Universit\"at Bochum,
  D-44780 Bochum, Germany}
\author{A.~Gasparyan}
\email[]{Email: ashotg@tp2.rub.de}
\affiliation{Institut f\"ur Theoretische Physik II, Ruhr-Universit\"at Bochum,
  D-44780 Bochum, Germany}
\affiliation{SSC RF ITEP, Bolshaya Cheremushkinskaya 25, 117218 Moscow, Russia}
\author{E.~Epelbaum}
\email[]{Email: evgeny.epelbaum@rub.de}
\affiliation{Institut f\"ur Theoretische Physik II, Ruhr-Universit\"at Bochum,
  D-44780 Bochum, Germany}
\date{\today}

\begin{abstract}
We derive the sub-subleading two-pion exchange contributions to the three-nucleon force which 
appear at next-to-next-to-next-to-next-to-leading order in chiral
effective field theory. In order to determine the low-energy constants, a complete analysis 
of pion-nucleon scattering at the subleading-loop order in the heavy-baryon expansion is 
carried out utilizing the power counting scheme employed in the derivation of the nuclear forces. 
We discuss the convergence of the chiral expansion for this particular
three-nucleon force topology and give the values of the low-energy
constants which provide the most realistic description of the
three-nucleon force when the chiral expansion is truncated at next-to-next-to-leading order. 
\end{abstract}


\maketitle

\vspace{-0.2cm}

\section{Introduction}
\def\theequation{\arabic{section}.\arabic{equation}}
\label{sec:intro}

\begin{figure}[tb]
\vskip 1 true cm
\includegraphics[width=0.9\textwidth,keepaspectratio,angle=0,clip]{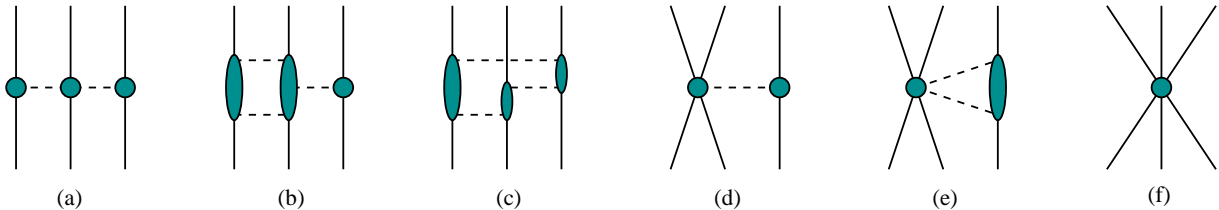}
    \caption{
          Various topologies contributing to the 3NF up to and
          including N$^4$LO:  two-pion ($2\pi$) exchange (a), two-pion-one-pion ($2\pi$-$1\pi$)
         exchange (b), ring (c), one-pion-exchange-contact (d),
         two-pion-exchange-contact (e) and purely contact (f) diagrams.
Solid and dashed lines
         represent nucleons and pions, respectively. 
Shaded blobs represent the corresponding amplitudes. 
\label{fig0} 
 }
\end{figure}

Three-nucleon forces (3NFs) are well known to play an important role
in nuclear physics.  In spite of many decades of effort, the detailed structure
of the 3NF is not properly reproduced by modern phenomenological 3NF
models, see Ref.~\cite{KalantarNayestanaki:2011wz}  for a
comprehensive review. This provides a strong motivation
to explore the structure of the 3NF within chiral effective field theory
(EFT) which is nowadays a standard tool to analyze low-energy nuclear
dynamics in harmony with the symmetries of QCD, see
Refs.~\cite{Epelbaum:2008ga,Machleidt:2011zz,Epelbaum:2012vx} for
recent  review articles.
In the past two decades, nuclear forces have already been extensively
studied in this framework. For two nucleons, it
turned out to be necessary and 
sufficient to go to next-to-next-to-next-to-leading order (N$^3$LO)  
in order to accurately describe the static
deuteron properties as well as the two-nucleon phase shifts and mixing
angles up to laboratory
energies of  $E_{\rm lab} \sim 200$ MeV  \cite{Entem:2003ft,Epelbaum:2004fk}. 
On the other hand, three- and more-nucleon
systems are so far only analyzed up to  next-to-next-to-leading order
(N$^2$LO) in the chiral expansion \cite{383846,nucl-th/0208023}. At this order one gets the first 
non-vanishing contributions to the 3NF which emerge
from the two-pion-exchange, one-pion-exchange-contact and purely
contact graphs (a), (d) and (f) in Fig.~\ref{fig0} with the
corresponding amplitudes being 
given by the lowest-order pion-nucleon 
vertices.
Generally, one observes a good description of nucleon-deuteron elastic and breakup scattering
observables at very low energies which improves when going from
next-to-leading order (NLO) to N$^2$LO.  On the other hand, the well-known
puzzles in the three-nucleon continuum such as e.g.~the $A_y$-puzzle \cite{Gloeckle:1995jg,Koike,Witala}
and the large discrepancy for the breakup cross section in the so-called
space-star and related configurations \cite{Duweke:2004xv,Ley:2006hu} still persist at
N$^2$LO. Notice, however, that  the recent calculation by the Pisa
group \cite{Viviani:2010mf} demonstrates that the $A_y$-puzzle in the 4N system is  
significantly reduced by the chiral 3NF at N$^2$LO.  The chiral EFT predictions for the
three-nucleon scattering observables at N$^2$LO at intermediate and higher
energies are, in general, in agreement 
with the data but show a rapidly increasing theoretical
uncertainty. It is, therefore,  necessary to go to higher orders in the chiral
expansion for three- and more-nucleon systems. 

The N$^3$LO contributions  to the 3NF emerge from the leading
relativistic corrections and pion-loop diagrams in all six
topologies shown in Fig.~\ref{fig0}. It is important to stress that
the N$^3$LO contributions   
do not involve any unknown low-energy constants (LECs). The
corresponding parameter-free expressions can be found in
Refs.~\cite{Bernard:2007sp,Bernard:2011zr}, see also
Ref.~\cite{Ishikawa:2007zz}. Another interesting feature of the
N$^3$LO 3NF corrections is their rather rich isospin-spin-momentum
structure emerging, especially, from the ring topology (c) in
Fig.~\ref{fig0}.  
This is in contrast with the quite restricted operator structure of the two-pion exchange 
3NF topology (a) whose effects in the three-nucleon continuum have already been extensively 
explored. It is, therefore, very interesting to study the impact of the novel structures 
in the 3NF on nucleon-deuteron scattering and the properties of light nuclei, especially in 
connection with the already mentioned unsolved puzzles. On the other hand, one may ask 
whether the resulting (leading) contributions to the structure functions accompanying the 
novel operator structures in the 3NF already allow for their decent description. Stated differently, 
the question is whether the lowest-nonvanishing-order contributions from the $2\pi$-$1\pi$ 
and ring-topologies are already converged or, at least, provide a reasonable approximation to 
the converged result. There is a strong reason to believe that this is not going to be 
the case since the contributions due to intermediate $\Delta$(1232) 
excitations are not yet taken into account for these topologies at N$^3$LO. 
In the standard chiral EFT formulation based 
on pions and nucleons as the only explicit degrees of freedom, effects of the $\Delta$
(and heavier resonances as well as heavy mesons) are hidden in the (renormalized) values of 
certain LECs starting from the subleading effective Lagrangian.  
The major part of the $\Delta$ contributions to the nuclear forces is taken into account 
in the $\Delta$-less theory through resonance saturation of the LECs $c_{3, 4}$ accompanying the 
subleading $\pi \pi NN$ vertices 
\cite{Ordonez:1993tn,Kaiser:1998wa,Krebs:2007rh,Epelbaum:2007sq,Epelbaum:2008td}  
(see, however, the last two references for some examples of the $\Delta$-contributions 
that go beyond the saturation of $c_{3,4}$). 
These LECs turn out to be numerically large and are known to be driven  
by the $\Delta$ isobar \cite{Bernard:1996gq,Krebs:2007rh}. As a consequence, one observes a 
rather unnatural convergence pattern in the chiral expansion of the two-pion exchange 
nucleon-nucleon potential $V_{\rm NN}^{2\pi}$ with by far the strongest contribution resulting from  
the formally subleading triangle diagram proportional to $c_3$ \cite{Kaiser:1997mw}.  
The (formally) leading contribution to $V_{\rm NN}^{2\pi}$ does not 
provide a good approximation to the potential so that one needs to go to (at least) the next-higher order 
in the chiral expansion and/or to include the $\Delta$ isobar as an explicit degree of freedom 
\cite{Krebs:2007rh}. The situation with the $2\pi$-$1\pi$ and ring topologies in the 3NF 
is similar. Based on the experience with the two-nucleon 
potential, one expects significant contributions due to intermediate  $\Delta$ excitations, 
see also the discussion in Ref.~\cite{Machleidt:2010kb}.  
For the ring topology, this expectation is confirmed by the phenomenological study of 
Ref.~\cite{Pieper:2001ap}. In order to include effects of the $\Delta$-isobar one needs 
\begin{itemize}  
\item
either to go to (at least) next-to-next-to-next-to-next-to-leading order (N$^4$LO) in the standard 
$\Delta$-less EFT approach
\item  
or to include the $\Delta$-isobar as an explicit degree of freedom.  
\end{itemize}
It should be understood that both strategies outlined above are, to some extent, complementary 
to each other. In particular, N$^3$LO contributions in the $\Delta$-less 
theory only take into account effects due to single $\Delta$-excitation but not due to the 
double and triple $\Delta$-excitations (whose inclusion in the $\Delta$-less approach would require  
the calculation at even higher orders).
These effects are taken into account already at N$^3$LO in the $\Delta$-full approach. On the other hand, 
there are also contributions not related to $\Delta$-excitations which are included/absent
in the $\Delta$-less approach at N$^4$LO/$\Delta$-full theory at N$^3$LO. It remains to be 
seen which strategy will turn out to be most efficient. The present paper represents the first step along this line. 
We analyze here the longest-range contribution to the 3NF in the standard, $\Delta$-less approach 
at N$^4$LO in the chiral expansion. This topology is particularly challenging due to (i) the need to carry 
out a non-trivial renormalization program as explained in section \ref{sec:TPE} and (ii) the need to re-consider 
pion-nucleon scattering in order to determine the relevant LECs. Our paper is organized as follows. In section 
\ref{sec:lagr}, we specify all terms in the effective Lagrangian that are needed in the calculation.  
The general structure of the two-pion exchange 3NF is discussed in section \ref{sec:TPE}. 
Here, we also briefly summarize the already  available results at N$^2$LO and N$^3$LO and give explicit 
expressions for the  N$^4$LO contributions. In section \ref{sec:piN} we analyze pion-nucleon 
scattering  at order $Q^4$ in the chiral expansion with $Q$ referring to the soft scale of the order 
of the pion mass  and use the available partial wave analyses to determine 
the relevant LECs. In section \ref{sec:results}, the numerical results for the two-pion exchange 3NF 
are presented and the convergence of the chiral expansion is discussed. Finally, the findings of 
our work are briefly summarized in section \ref{sec:summary}. The Appendices contain 
explicit formulae for the algebraic structure of the nuclear Hamiltonian at order N$^4$LO
and the chiral expansion of the $\pi N$ invariant amplitudes.

\section{Effective Lagrangian}
\def\theequation{\arabic{section}.\arabic{equation}}
\label{sec:lagr}

To derive the longest-range contributions to the 3NF at N$^4$LO we need
the effective pion-nucleon Lagrangian up to the order $Q^4$.  The
explicit expressions in the heavy-baryon formulation can be found in
Ref.~\cite{Bernard:1995dp,Fettes:2000gb}. 
For the sake of completeness, we list here all terms relevant for our
calculation with the corresponding building blocks being expanded in
powers of the pion fields:
\beqa
\mathcal{L}_{\pi N}^{(1)} &=&  N_v^\dagger \bigg[  i v \cdot \partial  - \frac{1}{4 F^2} \fet \tau \times \fet \pi \cdot (v \cdot \partial \fet \pi)  + 
\frac{8\alpha -1}{16 F^4}\fet \pi\cdot \fet \pi \,\fet \tau \times \fet \pi \cdot (v \cdot \partial \fet \pi)
    \nn
&& {} - \frac{\mathring{g}_A}{F} \fet \tau \cdot (S \cdot \partial \fet \pi )  
+ \frac{\mathring{g}_A}{2 F^3}  \left( (4 \alpha - 1) \fet \tau \cdot \fet \pi  \fet \pi \cdot (S \cdot \partial \fet \pi)  + 2 \alpha \fet \pi^2 \fet \tau \cdot (S\cdot \partial \fet \pi)   \right)  \bigg] N_v + \ldots      \,, \nn
\mathcal{L}_{\pi N}^{(2)} &=&   N_v^\dagger \bigg[ 4 M^2 c_1 - \frac{2}{F^2} c_1 M^2 \fet \pi^2  + \frac{1}{F^2} \left(c_2-\frac{g_A^2}{8m}\right) 
( v \cdot \partial \fet \pi) \cdot (v \cdot \partial\fet \pi ) + \frac{1}{F^2} c_3 (\partial _\mu \fet \pi ) \cdot (\partial^\mu \fet \pi )  \nn
&& {} - \frac{i}{F^2} \left(c_4+\frac{1}{4m}\right) \Big[ S_\mu , \; S_\nu \Big]  \fet \tau \times (\partial^\nu \fet \pi) \cdot (\partial^\mu \fet \pi ) 
+\frac{M^2 c_1}{2 F^4}(8\alpha -1)(\fet \pi\cdot \fet
\pi)^2+\frac{c_3}{F^4}\left((1-4\alpha)\fet \pi\cdot\partial_\mu\fet
  \pi\,\fet\pi\cdot\partial^\mu\fet\pi\right.\nn
&&{}\left. -2\,\alpha\,\fet \pi\cdot\fet \pi \,\partial_\mu\fet
  \pi\cdot\partial^\mu\fet \pi\right)-\frac{i\, c_4}{2
  F^4}\left(2\,(1-4\alpha)\fet \tau
  \cdot(\fet\pi\times\partial_\mu\fet\pi)\fet\pi\cdot\partial_\nu\fet
  \pi-4\,\alpha\, \fet \pi\cdot\fet \pi \,\partial_\mu\fet
  \pi\cdot(\fet\tau\times\partial_\nu\fet \pi)\right)\left[S^\mu,
  S^\nu\right] + \frac{\vec \nabla^2}{2m}\nn
&&{}+\frac{i \mathring g_A}{2 F m}\bigg(\fet \tau\cdot(v\cdot\partial\, S\cdot\partial\fet \pi)+2\fet \tau\cdot(v\cdot\partial\fet\pi)S\cdot\partial\bigg)+\frac{i}{8F^2 m}\bigg(\fet \tau\cdot(\fet \pi\times \fet(\vec\nabla^2\fet \pi))
+\fet\tau\cdot(\fet\pi\times\vec\nabla\fet\pi)\vec\nabla\bigg)
 \bigg]N_v+\ldots \,, \nn
\mathcal{L}_{\pi N}^{(3)} &=& N_v^\dagger \bigg[ \frac{2}{F^2} \bigg(d_1 + d_2-\frac{c_4}{4m}\bigg) \fet \tau \times (\partial_\mu v \cdot \partial \fet \pi) ) \cdot (\partial^\mu \fet \pi)  + \frac{2}{F^2} d_3 \fet \tau \times ( (v \cdot \partial)^2 \fet \pi ) \cdot (v \cdot \partial \fet \pi ) - \frac{4}{F^2} d_5 M^2 \fet \tau \times \fet \pi \cdot (v \cdot \partial \fet \pi ) \nn
&& {} - \frac{2 i }{F^2} (d_{14} - d_{15} )\Big[ (S \cdot \partial \, v \cdot \partial \fet \pi ), \; (S\cdot \partial \fet \pi )\Big]
-\frac{2}{F} (2d_{16} - d_{18}) M^2 \fet \tau \cdot (S \cdot \partial \, \fet \pi)\nn
&&{}+\frac{i\, c_2}{F^2 m}\bigg((v \cdot\partial\fet\pi)\cdot(\partial_\mu\fet\pi)\overrightarrow{\partial}^\mu-\overleftarrow{\partial}^\mu(v \cdot\partial\fet\pi)\cdot(\partial_\mu\fet\pi)\bigg)
-\frac{c_4}{2 F^2 m}\bigg(\fet \tau\cdot((v\cdot\partial\fet \pi)\times(\partial^2\fet\pi))\nn
&&{} -2\fet\tau\cdot((v\cdot\partial\fet\pi)\times(\partial_\mu\fet\pi))[S^\mu,S^\nu]\overrightarrow{\partial}_\nu+2\overleftarrow{\partial}_\nu\fet\tau\cdot((v\cdot\partial\fet\pi)\times(\partial_\mu\fet\pi))[S^\mu,S^\nu])\bigg)\bigg] N_v + \ldots \,, \nn
\mathcal{L}_{\pi N}^{(4)} &=& N_v^\dagger \bigg[  2 (8 e_{38} + e_{115} + e_{116} ) M^4  - \frac{i}{F^2}
\Big[ S_\mu , \; S_\nu \Big] \Big( - 18 e_{17} \fet \tau \times (\partial^\rho \partial^\mu \fet \pi ) \cdot 
 (\partial_\rho \partial^\nu \fet \pi ) 
 + e_{18} \fet \tau \times (v \cdot \partial \partial^\mu \fet \pi ) \cdot 
 (v \cdot \partial \partial^\nu \fet \pi )
\nn
&& {} + 4 (2 e_{21} - e_{37} )M^2 \fet \tau \times (\partial^\nu \fet \pi) \cdot (\partial^\mu \fet \pi ) \Big)
+ 8 e_{14} ( \partial_\mu \partial_\nu \fet \pi ) \cdot (\partial^\mu \partial^\nu \fet \pi)
+ 8 e_{15} ( v \cdot \partial \partial_\mu \fet \pi) \cdot ( v \cdot \partial \partial^\mu \fet \pi) \nn
&& {} + 8 e_{16} ( (v \cdot \partial )^2 \fet \pi) \cdot ( (v \cdot \partial )^2 \fet \pi) 
+ 4 M^2 (2 e_{19}- e_{22} - e_{36}) ( \partial_\mu \fet \pi) \cdot (\partial^\mu \fet \pi )
+ 8 e_{20} M^2  ( v \cdot \partial \fet \pi) \cdot (v \cdot \partial\fet \pi ) \nn
&& {} - 4 e_{22} M^2 \fet \pi \cdot ( \partial_\mu \partial^\mu \fet \pi ) 
- 8 e_{35} M^2 \fet \pi \cdot ((v \cdot \partial )^2 \fet \pi ) - 16 e_{38} M^4 \fet \pi^2 \bigg] N_v + \ldots \,,
\eeqa
where $\fet \pi$ and $N_v$ refer to pion and nucleon fields, $\fet \tau$ denote the isospin Pauli matrices, 
$v$ is the nucleon four-velocity and $S_\mu$ refers to  the covariant spin operator of the nucleon, 
\beq
S_\mu = \frac{1}{2} i \gamma_5 \sigma_{\mu \nu} v^\nu \,, \quad \quad
\sigma_{\mu \nu} = \frac{i}{2} [ \gamma_\mu , \; \gamma_\nu]\,. 
\eeq
Further, 
$F$ and $\mathring{g}_A$ are  the pion decay and the nucleon axial vector constants in the chiral limit, $M$ is the pion mass to leading order in quark masses  and 
$d_i $ ($e_i$) are further low-energy constants (LECs) from the order-$Q^3$ (order-$Q^4$) pion-nucleon Lagrangian.  
The superscript $i$ of $\mathcal{L}_{\pi N}^{(i)}$ refers to the number of
derivatives or insertions of the pion mass.  
Notice that in the power counting scheme employed
  here, the nucleon mass is treated as a heavier scale as compared to
  the chiral-symmetry-breaking scale $\Lambda_\chi$, see the
  discussion at the end of the section \ref{sec:TPE_2}. As a
  consequence,  the $1/m^2$-terms in $\mathcal{L}_{\pi N}^{(3)}$ and
the $1/m$-, $1/m^2$- and $1/m^3$-terms in $\mathcal{L}_{\pi
  N}^{(4)}$ generate contributions to the 3NF beyond N$^4$LO. We,
therefore, refrain from listing these terms in the effective Lagrangian. 
The constant $\alpha$ represents the freedom in parametrizing the pion field. Clearly, 
physical observables should not depend on a particular choice of $\alpha$. Therefore, keeping 
$\alpha$ unspecified and verifying $\alpha$-independence of the resulting nuclear forces 
allows for a non-trivial check of the calculation.  Notice that isospin-breaking corrections to the 3NF start 
to contribute at N$^3$LO. The complete expressions for isospin-breaking terms in the 3NF up to and including N$^4$LO 
can be found in Ref.~\cite{Epelbaum:2004xf}, see also \cite{Friar:2004rg,Friar:2004ca}. 
We, therefore, refrain from including isospin-breaking terms in the present work and employ exact 
isospin symmetry. For more details on the effective pion-nucleon Lagrangian the reader is referred to 
Refs.~\cite{Fettes:1998ud,Fettes:2000gb,Gasser:2002am}.

\section{The two-pion exchange 3NF}
\def\theequation{\arabic{section}.\arabic{equation}}
\label{sec:TPE}

The $2\pi$-exchange topology (a) generates the longest-range contribution
to the 3NF. In the isospin and static limits, its general structure in momentum space 
has the following form (modulo terms of a shorter range such as e.g.~the ones corresponding 
to the (b)-topology):
\beq
\label{2pi_general}
V_{2 \pi} = \frac{\vec \sigma_1 \cdot \vec q_1\,  \vec \sigma_3 \cdot
  \vec q_3}{[q_1^2 + M_\pi^2 ] \, [q_3^2 + M_\pi^2 ]}  \Big( \fet
  \tau_1 \cdot \fet \tau_3  \, {\cal A}(q_2) + \fet \tau_1 \times  \fet
\tau_3 \cdot \fet \tau_2  \,   \vec q_1 \times  \vec q_3   \cdot \vec
\sigma_2 \, {\cal B}(q_2) \Big)  \,,
\eeq
where $\vec \sigma_i$  denote the Pauli
spin matrices for the nucleon $i$ and $\vec q_{i} = \vec p_i \,
' - \vec p_i$,  with $\vec p_i \, '$ and $\vec p_i$ being the final and initial momenta of the nucleon $i$. 
Here and in what follows, we use the notation: $q_i \equiv | \vec q_i
|$.  The quantities ${\cal A} (q_2)$ and 
${\cal B} (q_2)$ in Eq.~(\ref{2pi_general}) are scalar
functions of the momentum transfer $q_2$ of the second nucleon whose explicite form is
derived within the chiral expansion. 
Unless stated otherwise, the expressions for the 3NF results are always
given for a particular choice of the nucleon labels. The complete result 
can then be found by taking into account all possible permutations of the
nucleons
\beq
V_{\rm 3N}^{\rm full} = V_{\rm 3N} + \mbox{5 permutations}\,.
\eeq

It is important to emphasize that we are using in Eq.~(\ref{2pi_general}) a slightly 
different notation as compared to e.g.~Ref.~\cite{Bernard:2007sp}. 
Specifically, in order to avoid the issue of non-uniqueness of
a decomposition into the $2\pi$, $2\pi$-$1\pi$ and shorter-range 
contributions, we get rid
of all terms which are proportional to $q_1^2$ and $q_3^2$. More precisely, using the
identity $q_{1,3}^2 = (q_{1,3}^2 + M_\pi^2) - M_\pi^2$  and canceling
the inverse pion propagator with the 
corresponding one in Eq.~(\ref{2pi_general}), each $q_{1,3}^2$ in the numerator gets 
replaced by $-M_\pi^2$ modulo  some additional contributions to the 
two-pion-one-pion exchange, one-pion-exchange-contact 
and the purely short-range contact interactions
$V_{\rm cont}$, see graphs (b), (d) and (f) in Fig.~\ref{fig0}.   
This way we ensure that the resulting functions $\cal A$ and $\cal B$ depend solely on $q_2$
rather than on $q_1$, $q_2$ and $q_3$.    
The purely short-range, induced terms $V_{\rm cont}$ only shift the low-energy constants
accompanying the contact 3NFs whose values anyway need to be 
adjusted to the data. There is no need to keep these contributions explicitly. 
The induced contributions to the other two topologies do, however, need to be 
taken into account, see Ref.~\cite{Friar:1998zt} for a related discussion.

\subsection{N$^2$LO and N$^3$LO contributions}
\label{sec:TPE_1}

We now briefly consider the first two terms in the chiral expansion of the functions ${\cal A} (q_2)$ and 
${\cal B} (q_2)$. 
The leading contributions arise at N$^2$LO which corresponds to 
the order $Q^3$ relative to the leading contribution to the nuclear
Hamiltonian
and have the form \cite{383846,nucl-th/0208023}
\beq
\label{ABQ2}
{\cal A}^{(3)} (q_2)= \frac{g_A^2}{8 F_\pi^4} \Big((2 c_3 - 4 c_1) M_\pi^2 +  c_3 
q_2^2  \Big) \,, \quad \quad
{\cal B}^{(3)} (q_2)= \frac{g_A^2 c_4}{8 F_\pi^4} \,, 
\eeq
where $g_A$, $F_\pi$ and $M_\pi$ denote to the physical values of the
nucleon axial vector coupling, pion decay constant and pion mass,
respectively, and the superscripts of  ${\cal A}$ and ${\cal B}$ refer to 
the powers of the soft scale $Q$.  The first corrections at N$^3$LO read
\cite{Bernard:2007sp,Ishikawa:2007zz}:
\beqa
\label{ABQ4}
{\cal A}^{(4)}(q_2) &=& \frac{g_A^4}{256 \pi  F_\pi^6} \Big[A(q_2)
\left(2 M_\pi^4+5 M_\pi^2 q_2^2+2 q_2^4
\right)+\left(4 g_A^2+1\right) M_\pi^3+2 \left(g_A^2+1\right) M_\pi 
q_2^2\Big]\,, \nn
{\cal B}^{(4)}(q_2) &=& -\frac{g_A^4 }{256 \pi  F_\pi^6} \Big[A(q_2) \left(4 M_\pi^2+q_2^2\right)+(2 g_A^2 
+1)M_\pi\Big]\,,
\eeqa
where the loop function $A(q)$ is defined as:
\beq 
A(q) = \frac{1}{2 q} \arctan \frac{q}{2 M_\pi}\,.
\eeq  
Notice that the leading-loop contributions to the $2\pi$-exchange
topology do not contain logarithmic ultraviolet divergences and, 
as explained in Ref.~\cite{Bernard:2007sp}, turn out to be
independent from the  LECs $d_i$ entering $\mathcal{L}_{\pi N}^{(3)}$.  
At both N$^2$LO and N$^3$LO, all LECs in the effective Lagrangian entering
the expressions  for the 3NF --
including $g_A$ and $F_\pi$ --  can be simply replaced by their
physical values.  

We further emphasize that, as already mentioned above,  the expressions 
in Eq.~(\ref{ABQ4}) differ from the ones in Eq.~(2.9) of Ref.~\cite{Bernard:2007sp}
by terms of a shorter range as compared to the two-pion exchange contributions. 
The advantage of using the new notation is that the results for $\cal A$ and $\cal B$ 
are now $\alpha$-independent.  This was not the case for terms in Eq.~(2.9) of 
Ref.~\cite{Bernard:2007sp} where the results are given for a specific choice $\alpha =0$.  

Last but not least, we emphasize that relativistic corrections to $V_{2 \pi}$ have a
richer structure than the one given in Eq.~(\ref{2pi_general}). The
explicit form of the $1/m$-corrections to   $V_{2 \pi}$ at N$^3$LO can be found in 
Ref.~\cite{Bernard:2011zr}, see also \cite{Friar:1994zz} for an early work.

\subsection{N$^4$LO contributions}
\label{sec:TPE_2}

\begin{figure}[tb]
\vskip 1 true cm
\includegraphics[width=15.0cm,keepaspectratio,angle=0,clip]{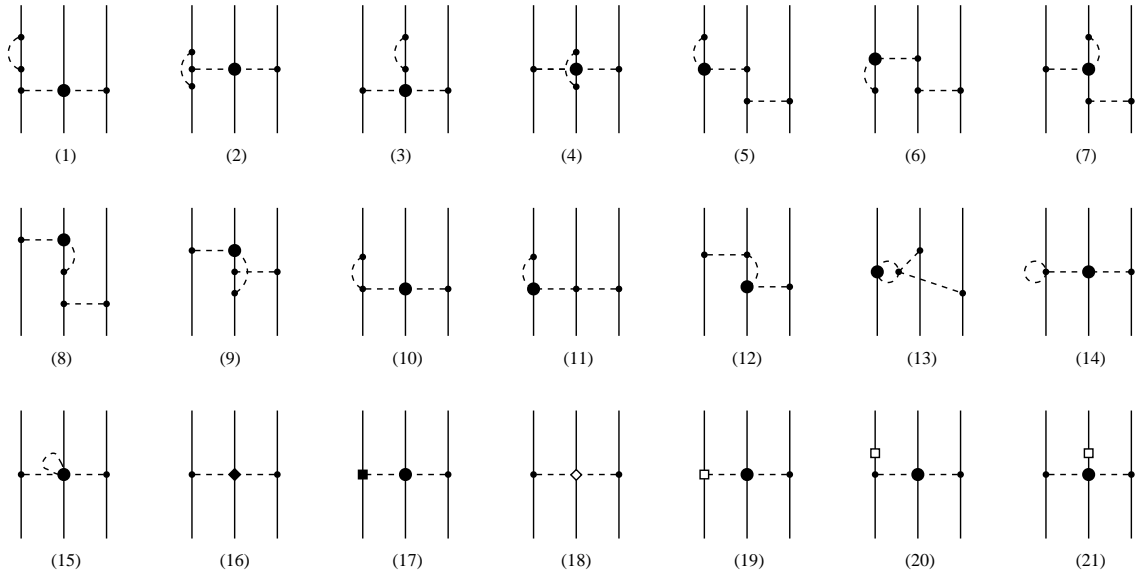}
    \caption{
         Two-pion exchange 3N diagrams at N$^4$LO. Solid dots, filled
 circles, rectangles and diamonds denote vertices from
$\mathcal{L}_{\pi N}^{(1)}$ or  $\mathcal{L}_{\pi}^{(2)}$, 
$\mathcal{L}_{\pi N}^{(2)}$, $\mathcal{L}_{\pi N}^{(3)}$ or
$\mathcal{L}_{\pi}^{(4)}$ and $\mathcal{L}_{\pi N}^{(4)}$,
respectively.  Open rectangles (diamonds) refer to $1/m$-vertices
from $\mathcal{L}_{\pi N}^{(2)}$ ($\mathcal{L}_{\pi N}^{(3)}$).  
 Diagrams which result from the interchange of the nucleon lines and/or
 application of the time reversal operation are not shown. For remaining
 notation see Fig.~\ref{fig0}.
\label{fig2} 
 }
\end{figure}

We now turn to the sub-subleading contributions to the $2\pi$-exchange 3NF  at order
$Q^5$ (N$^4$LO). These are depicted in Fig.~\ref{fig2} 
and emerge from: 
\begin{itemize} 
\item
one-loop diagrams (1)-(15) constructed from the leading-order vertices from
$\mathcal{L}_{\pi N}^{(1)}$ and a single  insertion of a subleading vertex $\propto c_i$, 
\item
tree diagram (16) involving leading-order vertices from
$\mathcal{L}_{\pi N}^{(1)}$ and a single  insertion of a vertex from $\mathcal{L}_{\pi N}^{(4)}$
proportional to LECs $e_i$,
\item
tree diagram (17) involving leading-order vertices from
$\mathcal{L}_{\pi N}^{(1)}$, one $\mathcal{L}_{\pi N}^{(2)}$-vertex
$\propto c_i$ and one  $\mathcal{L}_{\pi N}^{(3)}$-vertex
$\propto \bar d_i$,
\item 
relativistic $1/m$-corrections resulting from diagrams
(18)-(21).
\end{itemize}
Notice that since we work with renormalized pion field operators, we  
do not show explicitly in  Fig.~\ref{fig2} diagrams
involving pion self energy. We further emphasize that some of the diagrams
shown in Fig.~\ref{fig2}, such as e.g.~graphs (10) and (11), yield
vanishing contributions to the 3NF.  

It is important to keep in mind  that, in order to derive the genuine 3NF
contributions, we need to separate the irreducible parts in the
corresponding amplitudes to avoid double counting when iterating 
the potentials in the scattering equation. This is achieved employing the method of
unitary transformation, see Refs.~\cite{Epelbaum:1998ka,Epelbaum:2002gb}
for a comprehensive description of the method,  
Refs.~\cite{Epelbaum:2006eu,Epelbaum:2007us,Bernard:2007sp,Bernard:2011zr} for 
recent higher-order calculations of three- and four-nuclear forces and 
Refs.~\cite{Kolling:2009iq,Kolling:2011mt} for an extension to
electromagnetic processes. More precisely, we only use the method of
unitary transformation to evaluate the contributions of graphs (1),
(3), (5), (6)-(8), (20) and (21) in Fig.~\ref{fig0} which involve
reducible topologies. The remaining contributions are
obtained by calculating the corresponding Feynmann diagrams. 

Before showing our results for the N$^4$LO contributions to the functions 
$\cal A$ and $\cal B$, we briefly remind the reader how the calculations 
within the method of unitary transformation are organized. We begin with the 
effective Lagrangian specified in section \ref{sec:lagr} and 
switch to the Hamiltonian using the canonical formalism. We prefer to 
work with renormalized pion field $\fet \pi^r = Z_\pi^{-1/2} \fet \pi$ and mass $M_\pi$ 
and, therefore, do not need to deel with pion self-energy contributions.
In the second step, the pion degrees of
freedom are projected out by employing the appropriate unitary
transformation in the Fock space. The explicit form of the unitary
operator needed to compute nuclear forces up to N$^3$LO can be found in
Ref.~\cite{Epelbaum:2007us,Bernard:2011zr}. It is important to 
emphasize that getting rid of ultraviolet divergences appearing in the pion loop 
contributions to the nuclear Hamiltonian and current operators by expressing the bare 
LECs in terms of renormalized ones is not guaranteed a priori (given 
that the nuclear forces and currents are not observable quantities). 
This issue is discussed in detail
in Ref.~\cite{Epelbaum:2007us}, see also Ref.~\cite{Kolling:2011mt}. 
To ensure renormalizability, we exploit the unitary ambiguity
of the resulting nuclear potentials, i.e.~the freedom in choosing the
basis states in the Hilbert space. More precisely, we apply 
additional unitary transformations in the purely nucleonic 
subspace of the Fock space whose ``rotation angles'' are chosen 
in such a way, that all ultraviolet divergences are absorbed into 
redefinition of the LECs. As explicitly demonstrated in 
Refs.~\cite{Epelbaum:2007us,Kolling:2011mt}, it is possible to 
carry out this renormalization program for the leading-loop contributions 
to the 3NF and the electromagnetic two-nucleon current operators. 
An extension to subleading-loop 3NF contributions at N$^4$LO is straightforward. 
In Appendix \ref{app1} we show the resulting formal algebraic structure of the N$^4$LO 
contributions to the nuclear force $\propto g_A^4 c_i$ and the retardation corrections $\propto g_A^2
c_i/m$. 
To evaluate the corresponding 3NF contributions, we simply compute
the matrix elements of the connected time-ordered-like 3N diagrams
emerging from these operators and employ dimensional regularization to deal 
with ultraviolet divergencies. The unitary ambiguity is parametrized
via the  ``rotation angles'' $\alpha_{9,10,11}$ which have to be
chosen in such a way that the 3NF matrix elements become finite when
expressed in terms of renormalized LECs. At the order we are working, the 
relations between the LECs $\mathring{g}_A$, $F$ and $M$ and the corresponding 
renormalized constants are given by 
\beqa
\mathring{g}_A &=& g_A+\frac{g_A^3 M_\pi^2}{16 \pi^2 F_\pi^2}-4 d_{16}
M_\pi^2+\frac{2g_A(2g_A^2+1)\lambda_\pi
  M_\pi^2}{F^2}+\frac{g_A(c_3-2c_4)M_\pi^3}{6\pi F_\pi^2} + \mathcal{O}(M_\pi^4,
1/m)\,, \nn
F &=& F_\pi+\frac{M_\pi^2(2\lambda_\pi-l_4)}{F_\pi} + \mathcal{O}(M_\pi^4) \,, \nn
M^2 &=& M_\pi^2-\frac{M_\pi^4(2 l_3+\lambda_\pi)}{F_\pi^2} + \mathcal{O}(M_\pi^6)\,,
\eeqa
where $l_{3,4}$ are the LECs entering the subleading pion Lagrangian $\mathcal{L}_\pi^{(4)}$ 
\cite{Gasser:1983yg}
and the (divergent) quantity $\lambda_\pi$ is defined, following the notation of 
Ref.~\cite{Becher:2001hv},  via
\beq
\label{def_L}
  \lambda_\pi  =\frac{M^{d-4}}{16 \pi^2}\left[ \frac{1}{d-4} -
    \frac{1}{2}\left(\Gamma^\prime(1) + 1 + \log\left(4\pi \right) \right)\right].
\eeq
We also need the expression for the pion $Z$-factor $Z_\pi$, which has the form 
\beq
Z_\pi = 1 + \frac{2 M_\pi^2}{F_\pi^2} \left( (10 \alpha - 1) \lambda_\pi - l_4 \right) + \mathcal{O} 
 (M_\pi^4) \,.
\eeq
Notice that the $Z$-factor is not observable and shows an explicit dependence on $\alpha$. 
For $\alpha=0$, our result coincide with the one given in Ref.~\cite{Becher:2001hv}.    
Further, the LECs $l_i$, $d_i$ and $e_i$ can be decomposed into the
divergent parts and finite pieces. Utilizing the notation of Ref.~\cite{Becher:2001hv}\footnote{To simplify
the notation, we use here dimensionful $\epsilon_i$ in contrast to that reference.}  
this decomposition has the form:
 \beqa
  l_i  &=&  l_i^r (\mu) + \gamma_i \lambda = \frac{1}{16\pi^2}\bar{l}_i + \gamma_i \lambda_\pi \,, \nn
  d_i & = & d_i^r (\mu) + \frac{\delta_i}{F^2} \lambda = \bar{d}_i  +
  \frac{\delta_i}{F^2} \lambda_\pi \,, \nn
  e_i & = & e_i^r (\mu) + \frac{\epsilon_i}{F^2} \lambda = \bar{e}_i  +
  \frac{\epsilon_i}{F^2} \lambda_\pi \,,
\eeqa
where $\mu$ denotes the renormalization scale and 
the divergent, $\mu$-dependent quantity $\lambda$ is related to the $\mu$-independent one 
$\lambda_\pi$ through  
\beq
\lambda = \lambda_\pi - \frac{1}{32 \pi^2} \log \left( \frac{M^2}{\mu^2} \right)\,. 
\eeq
The coefficients $\gamma_i$ and $\delta_i$ are well known
in the framework of dimensional regularization adopted in the present work 
 \cite{Gasser:1983yg,Ecker:1995rk,Fettes:1998ud,Gasser:2002am}. The relevant coefficients 
read:
\beq
\delta_{18}  = 0, \quad \quad
 \delta_{16} = \frac{1}{2} \mathring{g}_A  + \mathring{g}_A^3,\quad  \quad \gamma_3 =
-\frac{1}{2}, \quad
 \quad  \gamma_4 =  2. 
\eeq
To the best of our knowledge, the $\epsilon_i$-coefficients have not yet been worked out. 
Fortunately, we only need a few linear combinations  of $\epsilon_i$ which are exactly 
the same as ones appearing in $\pi N$ scattering at order $Q^4$. Cancellation of 
the ultraviolet divergences in the $\pi N$ scattering amplitude implies the following relations 
for the divergent parts of the LECs $e_i$ 
\cite{Becher:2001hv}:
\beqa
\epsilon_{14} &=& -\frac{1}{12}c_2 -\frac{1}{2}c_3   \,, \nn
\epsilon_{17} &=&  -\frac{1}{12}c_4 \,, \nn
- 2 \epsilon_{19} + \epsilon_{22} + \epsilon_{36} &=& -2 c_1 +
\frac{5}{24} c_2 - \frac{3}{4} c_3   \,, \nn
- 2 \epsilon_{21} + \epsilon_{37} &=& -\frac{2}{3} c_4 - 3 g_A^2 c_4   \,, \nn
\epsilon_{22} - 4  \epsilon_{38} &=&  -3 c_1 + \frac{1}{4} c_2 +
\frac{3}{4} c_3   \,.
\eeqa
Expressing the N$^4$LO contribution to the 3NF in terms of $g_A$, $M_\pi$, $F_\pi$ and the
renormalized LECs $\bar l_i$, $\bar d_i$, $\bar e_i$ using the above relations leads 
to finite matrix elements provided the ``rotation angles'' $\alpha_{9,10,11}$  of the additional 
unitary transformations are chosen as:
\beq
\alpha_{10} = -\frac{1}{4}(1-2\alpha_9), \quad \quad
\alpha_{11} = \frac{1}{4}(1-2\alpha_9).
\eeq
Notice that while the parameter $\alpha_9$ is unfixed,  the resulting 3NF 
turns out to be $\alpha_9$-independent.  This leads to an unambiguous result for the 3NF 
at this order. We further emphasize that the obtained relations for $\alpha_{10,11}$  also constrain the form 
of the remaining N$^4$LO 3NF contributions, the four-nucleon force and 
the subleading-loop expressions for the exchange current operators.

The final, renormalized N$^4$LO contributions to the functions $\cal A$ and $\cal B$ 
in Eq.~(\ref{2pi_general}) have the form:
\beqa
{\cal A}^{(5)}(q_2)&=&\frac{g_A }{4608 \pi^2 F_\pi^6}\Big[M_\pi^2 q_2^2 \big(F_\pi^2 \left(2304 \pi^2 g_A (4 
\bar{e}_{14}+2 \bar{e}_{19}-\bar{e}_{22}-\bar{e}_{36})-2304 \pi^2 
\bar{d}_{18} c_3\right)\nn\
&+&g_A (144 c_1-53 c_2-90 c_3)\big)+M_\pi^4 
\left(F_\pi^2 \left(4608 \pi^2 \bar{d}_{18} (2 c_1-c_3)+4608 \pi^2 
g_A (2 \bar{e}_{14}+2 \bar{e}_{19}-\bar{e}_{36}-4 \bar{e}_{38})
\right)\right.\nn\
&+&\left. g_A \left(72 \left(64 \pi^2 \bar{l}_{3}+1\right) c_1-24 c_2-36 
c_3\right)\right)+q_2^4 \left(2304 \pi^2 \bar{e}_{14} F_\pi^2 g_A-2 
g_A (5 c_2+18 c_3)\right)\Big]\nn\
&-&\frac{g_A^2 }{768 \pi^2 F_\pi^6}
L(q_2) \left(M_\pi^2+2 q_2^2\right) \left(4 M_\pi^2 (6 c_1-c_2-3 
c_3)+q_2^2 (-c_2-6 c_3)\right) \,, \nn\
{\cal B}^{(5)}(q_2)&=&-\frac{g_A }{2304 \pi^2 F_\pi^6} \Big[M_\pi^2 \left(F_\pi^2 \left(1152 \pi^2 
\bar{d}_{18} c_4-1152 \pi^2 g_A (2 \bar{e}_{17}+2 
\bar{e}_{21}-\bar{e}_{37})\right)+108 g_A^3 c_4+24 g_A c_4
\right)\nn\
&+&q_2^2 \left(5 g_A c_4-1152 \pi^2 \bar{e}_{17} F_\pi^2 g_A
\right)\Big] + \frac{g_A^2 c_4 }{384 \pi^2 
F_\pi^6} L(q_2) \left(4 M_\pi^2+q_2^2\right)\,,
\eeqa
where the loop function $L (q)$ is defined according to  
\beq
L(q)  =  \frac{\sqrt{q^2 + 4 M_\pi^2}}{q} \log \frac{\sqrt{q^2 + 4
    M_\pi^2} + q}{2 M_\pi} \,.
\eeq
Interestingly, we observe that there are no $1/m$-contributions to
the two-pion exchange 3NF  at this order. In particular, no
N$^4$LO contributions emerge from diagrams (18) and (19) in Fig.~\ref{fig2}
since all leading (subleading) $1/m$-corrections to the $\pi NN$
($\pi\pi NN$)  vertex involve at least one time derivative. When
evaluating the corresponding Feynman diagrams, these time derivatives 
generate insertions of the nucleon kinetic energy which are further
suppressed by the factor $Q/m$. In addition, diagrms (20) and (21)
are found not to generate any irreducible pieces. 

The expressions  for the $2\pi$-exchange 3NF up to N$^4$LO discussed above 
depend on a number of low-energy constants. Here and in what follows, we use the 
values\footnote{Since
  we employ exact isospin limit in this work, we do not distinguish
  between the charge and neutral pion masses and use $M_\pi = 2
  M_{\pi^+}/3 + M_{\pi^0}/3$.}
\beq
g_A = 1.267 \,, \quad \quad 
F_\pi = 92.4 \mbox{ MeV}\,,   \quad \quad 
M_\pi = 138.03 \mbox{ MeV}\,. 
\eeq
The LECs $c_i$, $\bar d_i$ and $\bar e_i$ can be most
naturally determined from pion-nucleon scattering (at least) at the subleading-loop order 
(i.e.~$Q^4$). The heavy-baryon analyses of pion-nucleon scattering at orders $Q^3$ and 
$Q^4$ can be found in Refs.~\cite{Fettes:1998ud,Fettes:2000xg,Fettes:2001cr}, see also 
Refs.~\cite{Becher:2001hv,Alarcon:2011kh,Torikoshi:2002bt} for the calculations within the 
manifestly covariant framework, Ref.~\cite{Gasparyan:2010xz}  for a related calculation 
which extends chiral EFT to higher energies by employing constraints set 
by causality and unitarity and Ref.~\cite{Bernard:2007zu} for a recent review on baryon chiral 
perturbation theory. Unfortunately, we cannot use the values of the LECs obtained in 
these studies since we use a different counting scheme for the nucleon mass
in the few-nucleon sector, namely 
$Q/m \sim Q^2/\Lambda_\chi^2$ \cite{Weinberg:1991um}  rather then $m \sim \Lambda_\chi$ as used 
in the single-nucleon sector, see \cite{Epelbaum:2005pn} for an extended discussion.     
In the next section, we  re-analyze pion-nucleon scattering at order $Q^4$ in the 
heavy-baryon approach utilizing our counting scheme for the nucleon mass and
determine all relevant LECs from a fit to the available partial wave analyses.

\section{Determination of the LECs from $\pi N$ scattering at order $Q^4$}
\label{sec:piN}

In the center-of-mass system (cms), the amplitude for the reaction 
$\pi^a (q_1) + N(p_1) \to \pi^b(q_2) + N (p_2)$ with $p_{1,2}$ and
$q_{1,2}$ being the corresponding four-momenta and $a,b$ referring to
the pion isospin quantum numbers, takes the form:
\beq
T_{\pi N}^{ba} = \frac{E + m}{ 2 m }   \bigg( \delta^{ba} \Big[ g^+
(\omega, t) + i \vec \sigma \cdot \vec q_2 \times \vec q_1 \, h^+
(\omega, t) \Big]
+ i \epsilon^{bac} \tau^c \Big[ g^-
(\omega, t) + i \vec \sigma \cdot \vec q_2 \times \vec q_1 \, h^-
(\omega, t) \Big] \bigg)\,.
\eeq 
Here, $\omega = q_1^0 = q_2^0$ is the pion cms energy, 
$E_1 = E_2 \equiv E = ( \vec q \, ^2 + m^2 )^{1/2}$ the nucleon energy and
$\vec q_1 \, ^2 = \vec q_2 \, ^2 \equiv \vec q \, ^2 = ((s - M_\pi^2  -
m^2)^2  - 4 m^2 M_\pi^2)/(4s)$. 
Further, $t = (q_1  - q_2)^2$  is the invariant momentum transfer
squared while $s$ denotes the total  cms energy squared. The
quantities $g^\pm (\omega ,t)$ ($h^\pm (\omega ,t)$) refer to the isoscalar and isovector
non-spin-flip (spin-flip) amplitudes and  can be calculated  
in chiral perturbation theory. In Appendix \ref{app2}, we show the contributions to 
the amplitudes up to and including the order $Q^4$ using the same counting scheme 
for the nucleon mass as in the derivation of the nuclear forces.   
It is important to emphasize that terms in Eq.~(\ref{piNQ4})  proportional to 
the LECs $\bar {e}_{19}$, $\bar {e}_{20}$, $\bar {e}_{21}$, $\bar {e}_{22}$, $\bar {e}_{35}$, $\bar {e}_{36}$,  
$\bar {e}_{37}$, $\bar {e}_{38}$ and  $\bar{l}_3$ can be generated from the $Q^2$-terms by 
shifting the LECs $c_i$ as follows   \cite{Fettes:2000xg}:
\begin{eqnarray}
 c_1&\to& c_{1}-2M_{\pi}^{2}\left(\bar{e}_{22}-4\bar{e}_{38}-\frac{\bar{l}_3 c_1}{F_{\pi}^2}\right)\,,\nn
 c_2&\to& c_{2}+8M_{\pi}^{2}(\bar{e}_{20}+\bar{e}_{35})\,,\nn
 c_3&\to& c_{3}+4M_{\pi}^{2}\left(2\bar{e}_{19}-\bar{e}_{22}-\bar{e}_{36}+2\frac{\bar{l}_3 c_1}{F_{\pi}^2}\right)\,,\nn
 c_4&\to& c_{4}+4M_{\pi}^{2}(2\bar{e}_{21}-\bar{e}_{37})\,.
\end{eqnarray}
This implies that one cannot extract these combinations of $\bar{e}_i$ and $c_i$ 
separately from the $\pi N$ scattering data. We, therefore, follow Ref.~\cite{Fettes:2000xg} and absorb these 
linear combinations into redefinition of the $c_i$'s by setting 
\beq
\label{ei_conv}
\bar{e}_{22}-4\bar{e}_{38}-\frac{\bar{l}_3 c_1}{F_{\pi}^2} =0, \quad \quad
\bar{e}_{20}+\bar{e}_{35} =0 , \quad \quad
2\bar{e}_{19}-\bar{e}_{22}-\bar{e}_{36}+2\frac{\bar{l}_3 c_1}{F_{\pi}^2} =0, \quad \quad
2\bar{e}_{21}-\bar{e}_{37} =0\,,
\eeq
without loss of generality. Choosing another prescription would result in corrections 
generated by the loops involving $c_i$ that are beyond the order we are working.
Notice further that with the above convention, there is no dependence anymore on the LEC $\bar{l}_3$.
The LEC $\bar{d}_{18}$ can be fixed by means of the Goldberber-Treiman discrepancy
\begin{eqnarray}
g_{\pi N N}=\frac{g_{A}\, m}{F_{\pi}}\left(1-\frac{2M_{\pi}^{2}\,\bar{d}_{18}}{g_{A}}\right)\,,
\end{eqnarray}
where for $g_{\pi N N}$ we adopt the value from
Ref.~\cite{Timmermans:1990tz}: $g^2_{\pi N N}/(4\pi )\simeq 13.54$
which also agrees with the recent determination in Ref.~\cite{Baru:2010xn} 
based on the Goldberger-Miyazawa-Oehme sum rule and utilizing the 
most accurate available data on the pion-nucleon scattering lengths. 
Again, at the order we are working, we are free to set $\bar{d}_{18}=0$ provided 
we use the effective value for $g_A$, 
\beq
\label{ga_conv}
g_A = \frac{F_{\pi}g_{\pi N N}}{m} \simeq 1.285, 
\eeq
in all expressions. 
We adopt this convention in the following analysis. 
Thus, we are finally left with 13 independent (linear combinations of the) low energy constants 
to be fixed from a fit to the data, namely $c_{1,2,3,4}$, $\bar d_1 + \bar d_2$, $\bar d_3$, $\bar d_5$,
$\bar d_{14} - \bar d_{15}$ and $\bar e_{14,15,16,17,18}$.

The fit can be most conveniently performed in the partial wave basis using the available partial wave analyses.
In order to estimate a possible uncertainty of the extracted parameters, we considered two different
partial wave analyses in our fitting procedure, namely the one of  Ref.~\cite{Arndt:2006bf} by the 
group at the George Washington University, to be referred as GW, and the Karlsruhe-Helsinki analysis of   
Ref.~\cite{Koch:1985bn}), to be referred as KH.
The energy range of the data fitted corresponds to the $\pi N$ laboratory momenta 
$p_{\rm Lab}<150$ MeV/c. At higher energies the convergence of the chiral expansion becomes doubtful.
We follow the strategy which is similar to the one utilized in Ref.~\cite{Fettes:1998ud} and assign the 
same relative error to all empirical data points.

The partial wave amplitudes $f_{l\pm}^\pm (s)$, where $l$ refers to the
orbital angular momentum
and the subscript '$\pm$' to the total angular momentum ($j=l\pm s$), are given
in terms of the invariant amplitudes via
\begin{eqnarray*}
f_{l\pm}^\pm (s) = {E+m \over 16 \pi \sqrt{s}} \, \int_{-1}^{+1} dz \, \biggl[\,
g^\pm \, P_l (z) + \vec{q\,}^2 \, h^\pm \, (P_{l\pm1}(z) -zP_l (z) ) \biggr]~,
\end{eqnarray*}
where $z = \cos(\theta)$ is the angular variable ($t=2 \,\vec q \, ^2 (z-1)$). 
The amplitude in the isospin basis are related to  $f_{l\pm}^\pm$  as follows
\begin{eqnarray*}
 f_{l\pm}^{1/2}= f_{l\pm}^{+}+2f_{l\pm}^{-}\,,\quad \quad  f_{l\pm}^{3/2}= f_{l\pm}^{+}-f_{l\pm}^{-} \,.
\end{eqnarray*}
The phase shift for a partial wave amplitude with isospin $I$ is
obtained using the following unitarization 
prescription\footnote{It should be understood that this unitarization
  prescription goes, strictly speaking, beyond the chiral power
  counting. The resulting model dependence is, however,  very small
  due to the smallness of the phase shifts with the only exception of
  the $P_{33}$ partial wave, see \cite{Gasser:1990ku} for a related discussion.  
} \cite{Fettes:1998ud}:
\begin{eqnarray}
&& \delta^{I}_{l\pm}(s)=\arctan\Big(|\vec{q} \, |\,\Re \,f^{I}_{l\pm}(s)\Big)\,,
\end{eqnarray}
which reflects the absence of inelasticity below the two-pion production threshold.

\begin{figure}[tb]
\vskip 1 true cm
\includegraphics[width=15.0cm,keepaspectratio,angle=0,clip]{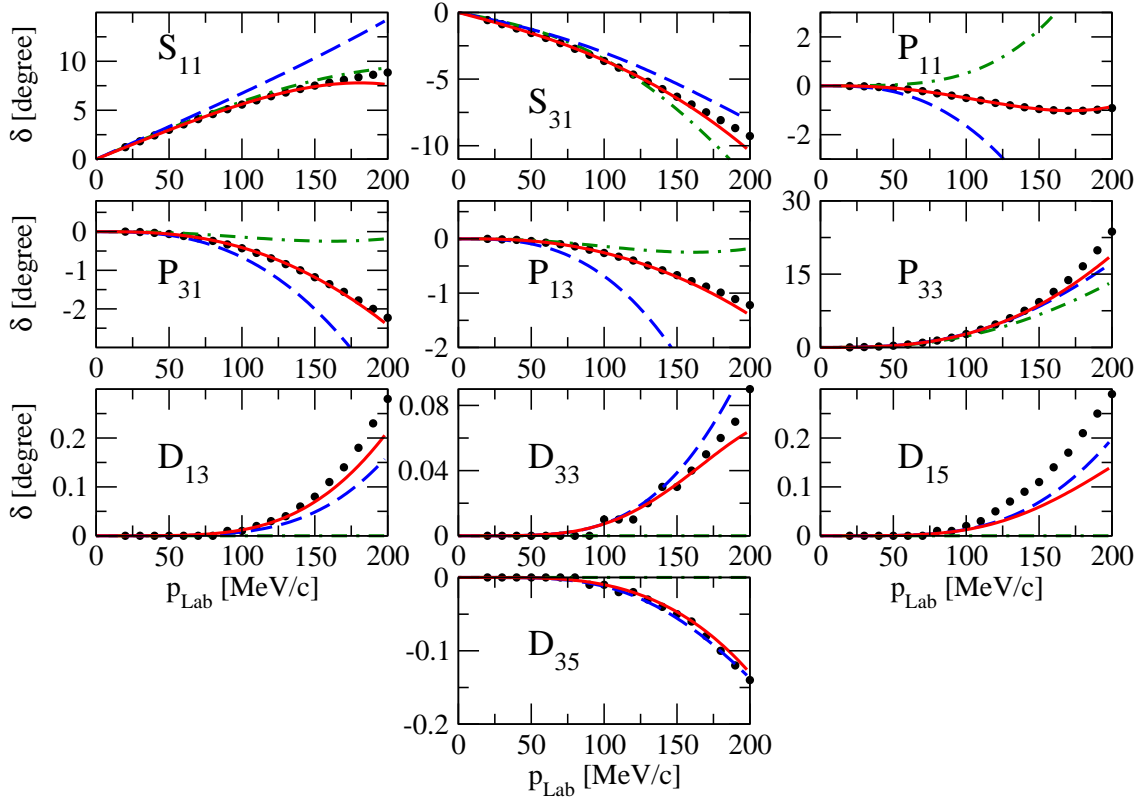}
\caption{Results of the fit for $\pi N$ $s$, $p$ and $d$-wave phase shifts using the GW partial wave analysis  of Ref.~\cite{Arndt:2006bf}.
The solid curves correspond to the full $Q^4$ results, the dashed curves
to the order-$Q^3$ results, and the dashed-dotted curves to the order-$Q^2$ calculation.         
 }\label{fig:SAID} 
\end{figure}

\begin{figure}[tb]
\vskip 1 true cm
\includegraphics[width=15.0cm,keepaspectratio,angle=0,clip]{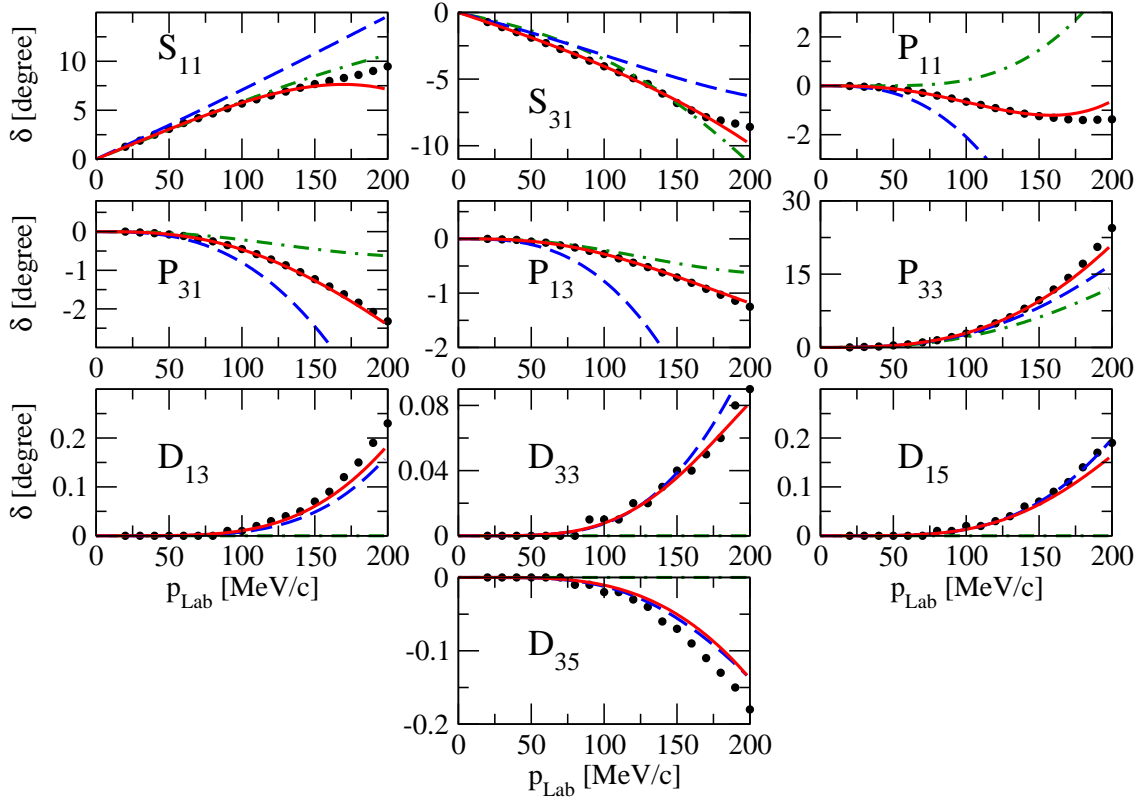}
\caption{Results of the fit for $\pi N$ $s$, $p$ and $d$-wave phase shifts using the KH partial wave analysis  of Ref.~\cite{Koch:1985bn}.
The solid curves correspond to the full $Q^4$ results, the dashed curves
to the order-$Q^3$ results, and the dashed-dotted curves to the order-$Q^2$ calculation.         
 }\label{fig:KA84}
\end{figure}

We performed a combined fit for all $s$-, $p$-, and $d$-waves since $d$-waves are the highest partial 
waves where the order-$Q^4$ counter terms contribute.
The results of the fits using the GW and KH partial wave analyses  
are visualized in Figs.~\ref{fig:SAID} and \ref{fig:KA84},
respectively. 
In these figures we show the full, order-$Q^4$ results (solid curves) as well 
as the phase shifts calculated up to the order $Q^3$ (dashed curves) and $Q^2$ (dashed-dotted curves) 
using the same parameters (from the order-$Q^4$ fit) in all curves. 
In the fitted region (from threshold up to $p_{\rm Lab}=150$ MeV/c), a good description of the 
data is achieved. As one would expect the convergence pattern when going from $Q^2$ to $Q^4$ 
is getting worse with increasing the pion momenta. Interestingly, the $d$-waves
are rather well reproduced already at the order $Q^3$ where there are no counter terms or other contributions 
depending on free parameters. Both the tree-level and finite loop contributions are important for those four 
partial waves. Our results for the phase shifts are similar and of a similar quality as the 
ones reported in Ref.~\cite{Fettes:2000xg}. 

We finally turn to the discussion of the extracted parameters. The obtained values of the 
low energy constants are collected in Table~\ref{table:parameters}.
\begin{table}[b]
\begin{center}
\begin{tabular}{|c|c|c|c|c|c|c|c|c|c|c|c|c|c|}
\hline 
 & $c_{1}$ & $c_{2}$ & $c_{3}$ & $c_{4}$ & $\bar{d}_{1}+\bar{d}_{2}$ & $\bar{d}_{3}$ & $\bar{d}_{5}$ & $\bar{d}_{14}-\bar{d}_{15}$ & $\bar{e}_{14}$ & $\bar{e}_{15}$ & $\bar{e}_{16}$ & $\bar{e}_{17}$ & $\bar{e}_{18}$\tabularnewline
\hline
fit to GW,  Ref.~\cite{Arndt:2006bf}&$-1.13$&$3.69$&$-5.51$&$3.71$&$5.57$&$-5.35$&$0.02$&$-10.26$&$1.75$&$-5.80$&$1.76$&$-0.58$&$0.96$\tabularnewline
\hline
 fit to KH,  Ref.~\cite{Koch:1985bn}&$-0.75$&$3.49$&$-4.77$&$3.34$&$6.21$&$-6.83$&$0.78$&$-12.02$&$1.52$&$-10.41$&$6.08$&$-0.37$&$3.26$\tabularnewline
\hline
\end{tabular}
\end{center}
\caption{Low-energy constants 
obtained from a fit to the empirical $s$, $p$- and $d$-wave pion-nucleon phase shifts using partial wave analysis  of Ref.~\cite{Arndt:2006bf}
and of  Ref.~\cite{Koch:1985bn}.
Values of the LECs are given in GeV$^{-1}$, GeV$^{-2}$ and  GeV$^{-3}$ for the $c_i$, $\bar{d}_i$ and $\bar{e}_i$, respectively.}
\label{table:parameters}
\end{table}
As one can see from the table, the LECs  $c_i$ and $\bar{d}_i$ turn out to come out rather similar 
for the two partial wave analyses. The difference does not exceed $30\%$ except 
for the LECs $c_1$ and $\bar d_5$ which are, however, considerably
smaller than the other $c_i$'s and $\bar d_i$'s, respectively.   The
same conclusion about stability can be drawn  for the LECs
$\bar{e}_{14}$ and $\bar{e}_{17}$. These are the only counter terms
contributing to $d$-waves, which is why these two constants are
strongly constrained by the threshold behavior of the $d$-wave phase
shifts. In contrast,  
the other $\bar{e}_i$'s 
are very sensitive to the energy dependence of the $s$- and $p$-wave
amplitudes and, therefore, vary strongly from one analysis to
another. Notice, however, that all extracted constants are of a
natural size 
except for the combination  $\bar{d}_{14}-\bar{d}_{15}$ and $\bar e_{15}$ which appear to be somewhat large.

We stress that one cannot directly compare the LECs $\bar{d}_i$ and $\bar{e}_i$ from of our fits to the ones 
obtained in Refs.~\cite{Fettes:1998ud},\cite{Fettes:2000xg} using heavy-baryon chiral perturbation theory 
at orders $Q^3$ and $Q^4$, respectively, because of a different power counting schemes in the two approaches. 
On the other hand, it is comforting to see that the extracted values for the 
$c_i$-, $\bar d_i$- and even some of the $\bar e_i$-coefficients are comparable to the 
ones found in Ref.~\cite{Fettes:2000xg} in the fit with the LECs $c_i$ being set to their 
order-$Q^3$ values, see table 4 of that work.   We also stress that the values for $c_{1,3,4}$ obtained 
from the fit to the KH partial wave analysis are in an excellent agreement with the 
ones  determined at order $Q^3$ by using chiral perturbation theory inside the  
 Mandelstam triangle \cite{Buettiker:1999ap}. It is also worth mentioning that the values of $c_{3,4}$ are in 
a good agreement with the ones determined from the new partial wave 
analysis of proton-proton and neutron-proton scattering data of Ref.~\cite{Rentmeester:2003mf}.  

It should be emphasized that one 
can obtain a considerably better description of the $\pi N$ phase shifts at orders $Q^2$ and $Q^3$ by 
allowing for the  LECs $c_i$ and $\bar d_i$ to be tuned rather than keeping their values fixed at order $Q^4$. 
In fact, the values of $c_i$ are well known to change significantly when performing 
fits at orders $Q^2$ and $Q^3$. Using the KH partial wave analysis, employing the 
order-$Q^2$ expressions for the amplitudes and utilizing the
same fitting procedure as before, we end up with the following values for the $c_i$'s:
\beq
\label{ci_fitQ2}
c_1^{\rm KH} = -0.26 \mbox{ GeV}^{-1}, \quad \quad
c_2^{\rm KH}  = 2.02 \mbox{ GeV}^{-1}, \quad \quad
c_3^{\rm KH}  = -2.80 \mbox{ GeV}^{-1}, \quad \quad
c_4^{\rm KH}  = 2.01 \mbox{ GeV}^{-1}\,,
\eeq
while the GW partial wave analysis yields:
\beq
\label{ci_fitQ2_GW}
c_1^{\rm GW} = -0.58 \mbox{ GeV}^{-1}, \quad \quad
c_2^{\rm GW}  = 2.02 \mbox{ GeV}^{-1}, \quad \quad
c_3^{\rm GW}  = -3.14 \mbox{ GeV}^{-1}, \quad \quad
c_4^{\rm GW}  = 2.19 \mbox{ GeV}^{-1}\,.
\eeq
Notice that $c_{2,3,4}$ turn out to be somewhat smaller in magnitude than the ones extracted from 
the order-$Q^2$ fit to the $s$- and $p$-wave $\pi N$ threshold coefficients 
\cite{Krebs:2007rh}.\footnote{This  indicates that the order-$Q^2$ representation of the amplitudes does not provide 
the appropriate desciption of the data in the whole momentum range used in our fits.}  
We will come back to the issue of optimizing the description of the data at 
lower orders of the chiral expansion in the next section. Notice, however, that in spite of such a possibility, 
we believe that the results shown in Figs.~\ref{fig:SAID}    
and \ref{fig:KA84} provide a more realistic picture of the convergence of the heavy-baryon 
chiral perturbation theory for pion-nucleon scattering.

\section{Results for the two-pion
  exchange 3NF}
\label{sec:results}

\begin{figure}[tb]
\vskip 1 true cm
\includegraphics[width=0.8\textwidth,keepaspectratio,angle=0,clip]{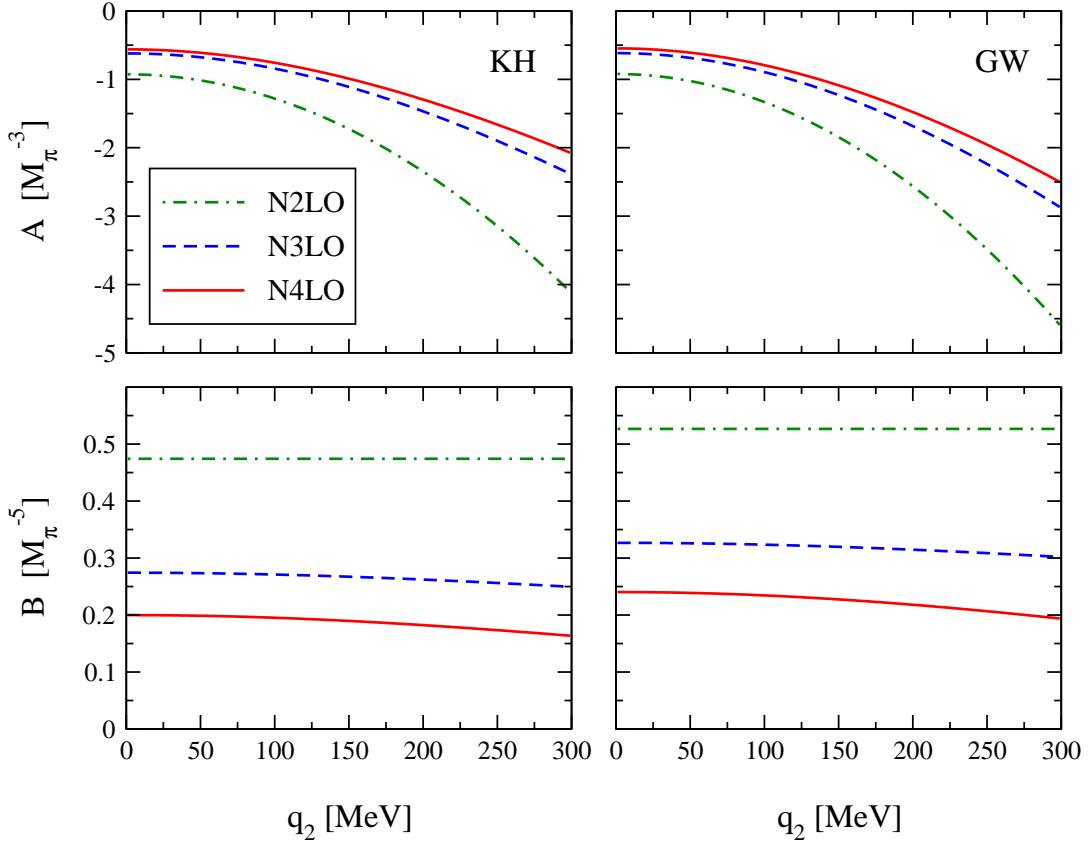}
    \caption{Chiral expansion of the functions ${\cal A} (q_2)$ and ${\cal B} (q_2)$
entering the two-pion exchange 3NF in Eq.~(\ref{2pi_general}) up to N$^4$LO.  Left (right)
panel shows the results obtained with the LECs determined from the order-$Q^4$ fit to 
the pion-nucleon partial wave analysis of Ref.~\cite{Koch:1985bn} (Ref.~\cite{Arndt:2006bf}) and listed 
in Table \ref{table:parameters}. 
Dashed, dashed-dotted and solid lines correspond to 
${\cal A}^{(3)}$,  ${\cal A}^{(3)} +  {\cal A}^{(4)}$ and 
${\cal A}^{(3)} +  {\cal A}^{(4)} +  {\cal A}^{(5)}$  in the upper plots while
${\cal B}^{(3)}$,  ${\cal B}^{(3)} +  {\cal B}^{(4)}$ and 
${\cal B}^{(3)} +  {\cal B}^{(4)} +  {\cal B}^{(5)}$ in the lower plots. 
\label{fig:res} 
 }
\end{figure}

With all relevant LECs being determined from pion-nucleon 
scattering, we are now in the position to analyze the convergence of the 
chiral expansion for the two-pion exchange 3NF. In Fig.~\ref{fig:res}, 
we show the results for the functions ${\cal A} (q_2)$ and ${\cal B} (q_2)$
for small values of the momentum transfer $q_2$, $q_2 < 300$ MeV 
at various orders in the chiral expansion. 
More precisely, we plot ${\cal A}^{(3)}$,  ${\cal A}^{(3)} +  {\cal A}^{(4)}$ and 
${\cal A}^{(3)} +  {\cal A}^{(4)} +  {\cal A}^{(5)}$ as well as 
${\cal B}^{(3)}$,  ${\cal B}^{(3)} +  {\cal B}^{(4)}$ and 
${\cal B}^{(3)} +  {\cal B}^{(4)} +  {\cal B}^{(5)}$ using the values 
of the LECs $c_i$, $\bar d_i$ and $\bar e_i$  determined from 
the order-$Q^4$ fit to the KH and GW partial wave analyses 
as described in the previous section. We use here the same, fixed values for the 
LECs $c_i$ (and $\bar d_i$)  listed in Table \ref{table:parameters} at all orders 
and adopt the same conventions regarding the LECs as in the case of pion-nucleon 
scattering, see  Eqs.~(\ref{ei_conv}) and (\ref{ga_conv}). Notice that
${\cal A}^{(5)}$ and ${\cal B}^{(5)}$ do not depend on the LECs $\bar
e_{15,16,18}$ which are very sensitive to a particular  choice of the partial wave  
analysis in pion-nucleon scattering, see Table \ref{table:parameters}.  
The relevant LECs $\bar e_{14,17}$ are, on the contrary,  rather
stable as they are well determined in $\pi N$ $d$-waves.  

One observes a very good convergence for the function $\cal A$ with 
the subleading-order result (i.e. N$^3$LO) being very close to the 
one at N$^4$LO. It is also comforting to see that both partial wave analyses 
lead to similar results for this quantity. The dependence on the input 
for pion-nucleon phase shifts for $\cal A$ is bigger than the shift 
from N$^3$LO to N$^4$LO which can serve as a (conservative) estimation 
of the theoretical uncertainty at N$^4$LO. The convergence for the 
function $\cal B$ is somewhat slower with the shift from N$^3$LO to N$^4$LO
being of the order of $\sim 30$\%. Also the difference between the two 
partial wave analyses of the order of $\sim 20$\%  is larger than 
for the function $\cal A$.  
It should be understood that an accurate description of the low-energy pion-nucleon 
scattering data at different orders does not automatically guarantee a good 
convergence of the chiral expansion for $\cal A$ and $\cal B$. In particular, 
these quantities do not depend on the LECs $\bar d_i$ (to the order considered)
which contribute to $\pi N$ phase shifts. Thus, the observed   
reasonable convergence for the $2\pi$-exchange 3NF is a highly 
non-trivial test of the theoretical approach.

\begin{figure}[tb]
\vskip 1 true cm
\includegraphics[width=0.8\textwidth,keepaspectratio,angle=0,clip]{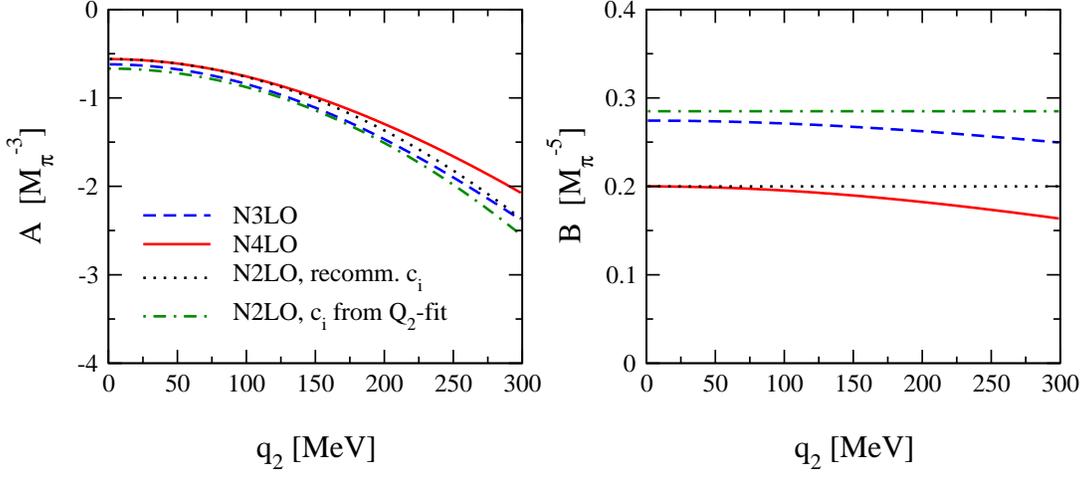}
    \caption{Chiral expansion of the functions ${\cal A} (q_2)$ and ${\cal B} (q_2)$ using the LECs 
determined from the fits to the KH $\pi N$ partial wave analysis. The dashed and solid lines are 
the same as in Fig.~\ref{fig:res}. The dotted and dashed-dotted lines show ${\cal A}^{(3)} (q_2)$,   ${\cal B}^{(3)} (q_2)$ 
with the LECs $c_i$ taken from Eq.~(\ref{ci_rec}) and (\ref{ci_fitQ2}), respectively.
\label{fig:res_recomm} 
 }
\end{figure}

Given that the $q_2$-dependence of  $\cal A$, $\cal B$ does not change 
significantly when going from N$^2$LO to N$^4$LO, it is 
clear that the final result at N$^4$LO can be well approximated 
by the N$^2$LO expressions with the appropriately tuned LECs
$c_{1,3,4}$. This feature is visualized in Fig.~\ref{fig:res_recomm},  
where the dotted lines show ${\cal A}^{(3)} (q_2 )$  and ${\cal B}^{(3)} (q_2 )$ from  Eq.~(\ref{ABQ2})  
with   
\beq
\label{ci_rec}
c_1^{\rm KH} = -0.37 \mbox{ GeV}^{-1}, \quad \quad
c_3^{\rm KH} = -2.71 \mbox{ GeV}^{-1}, \quad \quad
c_4^{\rm KH} = 1.41 \mbox{ GeV}^{-1}\,.
\eeq
These values of $c_i$'s are adjusted in such a way that  ${\cal A}^{(3)} (0)$,   ${\cal B}^{(3)} (0)$ and the curvature 
of the function   ${\cal A}^{(3)} (q_2 )$ at $q_2 =0$  coincide with the ones resulting from 
${\cal A}^{(3)} +  {\cal A}^{(4)} +  {\cal A}^{(5)}$, ${\cal B}^{(3)} +  {\cal B}^{(4)} +  {\cal B}^{(5)}$
and using the LECs from the KH fits to the $\pi N$ data.
Not surprisingly, the $c_i$'s in Eq.~(\ref{ci_rec}) 
are fairly close to the $Q^2$-fit values in Eq.~(\ref{ci_fitQ2}). We also show in Fig.~\ref{fig:res_recomm} 
 the functions ${\cal A}^{(3)} (q_2)$,   ${\cal B}^{(3)} (q_2)$ with the $c_i$'s
from Eq.~(\ref{ci_fitQ2}) (dashed-dotted lines).  When using the GW
partial wave analysis, the resulting values of the $c_i$'s slightly
change to:
\beq
\label{ci_rec_2}
c_1^{\rm GW} = -0.73 \mbox{ GeV}^{-1}, \quad \quad
c_3^{\rm GW} = -3.38 \mbox{ GeV}^{-1}, \quad \quad
c_4^{\rm GW} = 1.69 \mbox{ GeV}^{-1}\,.
\eeq
The values  in Eqs.~(\ref{ci_rec}) and (\ref{ci_rec_2})  can be regarded as the recommended ones when the 3NF 
is taken into account at N$^2$LO. One should, however, always keep in mind that the true 
theoretical uncertainty for the $2\pi$-exchange 3NF at N$^2$LO is sizable, see Fig.~\ref{fig:res}.

\section{Summary and conclusions}
\def\theequation{\arabic{section}.\arabic{equation}}
\label{sec:summary}

In this paper, we have analyzed the longest-range contribution to the three-nucleon force 
at N$^4$LO utilizing the heavy-baryon formulation of chiral EFT with pions and nucleons 
as the only explicit degrees of freedom. For this particular topology, the N$^4$LO corrections 
already provide the sub-subleading contribution, so that one can address the question 
of convergence of the chiral expansion. The pertinent results of our study can be summarized 
as follows. 
\begin{itemize}
\item
We worked out the N$^4$LO contributions to the $2\pi$-exchange 3NF. The unitary ambiguity 
of the Hamilton operator can be parametrized at this order by three additional unitary transformations. 
We found that two of the corresponding ``rotation angles'', namely $\alpha_{10}$ and $\alpha_{11}$, 
are fixed in terms of the remaining one $\alpha_9$ if one requires that the resulting 
3NF matrix elements are finite (renormalizability constraint). The parameter $\alpha_9$
does not enter the expressions for the 3NF at N$^4$LO. These findings will impact the 
results for the remaining 3NF contributions which are not considered in this paper. 
\item
In order to determine the low-energy constants $c_i$, $\bar d_i$ and $\bar e_i$ 
contributing to the $2\pi$-exchange  3NF, we re-analyzed pion-nucleon scattering 
at order $Q^4$ employing exactly the same power counting scheme as in the derivation 
of the nuclear forces.  We used the available partial wave analyses of the pion-nucleon 
scattering data to determine all relevant LECs. The resulting values turn out to be 
rather stable and agree well with the determinations by other groups.  
\item 
With all LECs being fixed from pion-nucleon scattering as discussed 
above, we found a good/reasonable convergence of the chiral expansion for 
the functions $\cal A$ and $\cal B$ which parametrize the (static part of the) 
two-pion exchange 3NF. We also provide the recommended values of the 
LECs $c_{1,3,4}$ which allow one to approximate the full, N$^4$LO result for the 
$2\pi$-exchange 3NF by the N$^2$LO expressions.   
\end{itemize}

As explained in the introduction, one generally expects important contributions to 
the nuclear forces associated with the intermediate $\Delta$-excitations.  
The observed (reasonable) convergence pattern for the longest-range 3NF 
is not surprising given that effects of the $\Delta$-isobar are, to a large extent, 
accounted for already in the leading contribution to this topology (i.e.~at N$^2$LO)
through resonance saturation of the LECs $c_{3,4}$. The situation is different for the 
two-pion-one-pion exchange and ring 3NF topologies, whose leading contributions 
at N$^3$LO are completely missing effects of the $\Delta$-isobar.  The 
corresponding N$^4$LO corrections are, therefore, expected to be large and need to 
be worked out. Work along these lines is in progress.

\section*{Acknowledgments}

We are grateful to Ulf-G.~Mei{\ss}ner for helpful comments on the
manuscript and Achim Schwenk for 
useful discussions. 
This work is supported by the EU HadronPhysics3 project ``Study of strongly interacting matter'', 
by the European Research Council (ERC-2010-StG 259218 NuclearEFT) and 
by the DFG (TR 16, ``Subnuclear Structure of Matter'').

\appendix

\def\theequation{\Alph{section}.\arabic{equation}}
\setcounter{equation}{0}
\section{Formal algebraic structure of the N$^4$LO corrections}
\label{app1}

In this Appendix we list the formal
operator structure of the various N$^4$LO contributions to the nuclear
Hamiltonian relevant for the present calculations. A detailed
discussion on the method of unitary transformation including the explicit form of the 
unitary operator at lower orders in the chiral expansion can be 
found in Ref.~\cite{Epelbaum:2007us}.
\begin{itemize}
\item{terms $\propto g_A^4 c_i$:}
\beqa
V&=& \eta \bigg[   \alpha_9 \bigg(
  H_{21}^{(1)}  \frac{\lambda^1}{E_\pi}  H_{21}^{(1)}  \eta  \
H_{21}^{(1)}  \frac{\lambda^1}{E_\pi}  H_{21}^{(1)}  \
\frac{\lambda^2}{E_\pi^2}  H_{22}^{(3)}  
 -   H_{21}^{(1)}  \
\frac{\lambda^1}{E_\pi}  H_{21}^{(1)}  \eta  H_{22}^{(3)}  \
\frac{\lambda^2}{E_\pi^2}  H_{21}^{(1)}  \frac{\lambda^1}{E_\pi}  \
H_{21}^{(1)}\nn
&&{}   -  H_{21}^{(1)}  \frac{\lambda^1}{E_\pi}  \
H_{21}^{(1)}  \frac{\lambda^2}{E_\pi^2}  H_{22}^{(3)}  \eta  \
H_{21}^{(1)}  \frac{\lambda^1}{E_\pi}  H_{21}^{(1)}   +   \
H_{22}^{(3)}  \frac{\lambda^2}{E_\pi^2}  H_{21}^{(1)}  \
\frac{\lambda^1}{E_\pi}  H_{21}^{(1)}  \eta  H_{21}^{(1)}  \
\frac{\lambda^1}{E_\pi}  H_{21}^{(1)}  \bigg)\nn
&& {} + 
\alpha_{10} \bigg(   H_{21}^{(1)}  \frac{\lambda^1}{E_\pi}  \
H_{21}^{(1)}  \eta  H_{21}^{(1)}  \frac{\lambda^1}{E_\pi^2}  \
H_{21}^{(1)}  \frac{\lambda^2}{E_\pi}  H_{22}^{(3)}  -   \
H_{21}^{(1)}  \frac{\lambda^1}{E_\pi}  H_{21}^{(1)}  \eta  \
H_{22}^{(3)}  \frac{\lambda^2}{E_\pi}  H_{21}^{(1)}  \
\frac{\lambda^1}{E_\pi^2}  H_{21}^{(1)}  \nn
&& {}  -  H_{21}^{(1)}  \
\frac{\lambda^1}{E_\pi^2}  H_{21}^{(1)}  \frac{\lambda^2}{E_\pi}  \
H_{22}^{(3)}  \eta  H_{21}^{(1)}  \frac{\lambda^1}{E_\pi}  \
H_{21}^{(1)}   +   H_{22}^{(3)}  \frac{\lambda^2}{E_\pi}  \
H_{21}^{(1)}  \frac{\lambda^1}{E_\pi^2}  H_{21}^{(1)}  \eta  \
H_{21}^{(1)}  \frac{\lambda^1}{E_\pi}  H_{21}^{(1)}  \bigg) \nn
&& {}+
\alpha_{11} \bigg( H_{21}^{(1)}  \frac{\lambda^1}{E_\pi}  \
H_{21}^{(1)}  \eta  H_{21}^{(1)}  \frac{\lambda^1}{E_\pi}  \
H_{22}^{(3)}  \frac{\lambda^1}{E_\pi^2}  H_{21}^{(1)}   - 
H_{21}^{(1)}  \frac{\lambda^1}{E_\pi}  H_{21}^{(1)}  \eta  \
H_{21}^{(1)}  \frac{\lambda^1}{E_\pi^2}  H_{22}^{(3)}  \
\frac{\lambda^1}{E_\pi}  H_{21}^{(1)}  \nn
&& {}  -  H_{21}^{(1)}  \
\frac{\lambda^1}{E_\pi}  H_{22}^{(3)}  \frac{\lambda^1}{E_\pi^2}  \
H_{21}^{(1)}  \eta  H_{21}^{(1)}  \frac{\lambda^1}{E_\pi}  \
H_{21}^{(1)}   +  H_{21}^{(1)}  \frac{\lambda^1}{E_\pi^2}  \
H_{22}^{(3)}  \frac{\lambda^1}{E_\pi}  H_{21}^{(1)}  \eta  \
H_{21}^{(1)}  \frac{\lambda^1}{E_\pi}  H_{21}^{(1)}  \bigg) \nn
&& {}
+ \frac{1}{2} \bigg( -   H_{21}^{(1)}  \frac{\lambda^1}{E_\pi}  H_{21}^{(1)}  \eta  \
H_{21}^{(1)}  \frac{\lambda^1}{E_\pi}  H_{21}^{(1)}  \
\frac{\lambda^2}{E_\pi^2}  H_{22}^{(3)}   -   H_{21}^{(1)}  \
\frac{\lambda^1}{E_\pi}  H_{21}^{(1)}  \eta  H_{21}^{(1)}  \
\frac{\lambda^1}{E_\pi}  H_{22}^{(3)}  \frac{\lambda^1}{E_\pi^2}  \
H_{21}^{(1)}  \nn
&& {} -   H_{21}^{(1)}  \frac{\lambda^1}{E_\pi}  \
H_{21}^{(1)}  \eta  H_{21}^{(1)}  \frac{\lambda^1}{E_\pi^2}  \
H_{21}^{(1)}  \frac{\lambda^2}{E_\pi}  H_{22}^{(3)}   -   \
H_{21}^{(1)}  \frac{\lambda^1}{E_\pi}  H_{21}^{(1)}  \eta  \
H_{21}^{(1)}  \frac{\lambda^1}{E_\pi^2}  H_{22}^{(3)}  \
\frac{\lambda^1}{E_\pi}  H_{21}^{(1)}  \nn
&&  {} -   H_{21}^{(1)}  \
\frac{\lambda^1}{E_\pi}  H_{21}^{(1)}  \eta  H_{22}^{(3)}  \
\frac{\lambda^2}{E_\pi}  H_{21}^{(1)}  \frac{\lambda^1}{E_\pi^2}  \
H_{21}^{(1)}   -   H_{21}^{(1)}  \frac{\lambda^1}{E_\pi}  \
H_{21}^{(1)}  \eta  H_{22}^{(3)}  \frac{\lambda^2}{E_\pi^2}  \
H_{21}^{(1)}  \frac{\lambda^1}{E_\pi}  H_{21}^{(1)}  \nn
&& {}  + 2   \,
H_{21}^{(1)}  \frac{\lambda^1}{E_\pi}  H_{21}^{(1)}  \
\frac{\lambda^2}{E_\pi}  H_{21}^{(1)}  \frac{\lambda^1}{E_\pi}  \
H_{21}^{(1)}  \frac{\lambda^2}{E_\pi}  H_{22}^{(3)}   + 2   \,
H_{21}^{(1)}  \frac{\lambda^1}{E_\pi}  H_{21}^{(1)}  \
\frac{\lambda^2}{E_\pi}  H_{21}^{(1)}  \frac{\lambda^1}{E_\pi}  \
H_{22}^{(3)}  \frac{\lambda^1}{E_\pi}  H_{21}^{(1)}  \nn
&& {}  + 2  \,
H_{21}^{(1)}  \frac{\lambda^1}{E_\pi}  H_{21}^{(1)}  \
\frac{\lambda^2}{E_\pi}  H_{21}^{(1)}  \frac{\lambda^3}{E_\pi}  \
H_{21}^{(1)}  \frac{\lambda^2}{E_\pi}  H_{22}^{(3)}   + 2  \,
H_{21}^{(1)}  \frac{\lambda^1}{E_\pi}  H_{21}^{(1)}  \
\frac{\lambda^2}{E_\pi}  H_{21}^{(1)}  \frac{\lambda^3}{E_\pi}  \
H_{22}^{(3)}  \frac{\lambda^1}{E_\pi}  H_{21}^{(1)}  \nn
&& {}  -  \
H_{21}^{(1)}  \frac{\lambda^1}{E_\pi}  H_{21}^{(1)}  \
\frac{\lambda^2}{E_\pi}  H_{22}^{(3)}  \eta  H_{21}^{(1)}  \
\frac{\lambda^1}{E_\pi^2}  H_{21}^{(1)}   + 2  \, H_{21}^{(1)}  \
\frac{\lambda^1}{E_\pi}  H_{21}^{(1)}  \frac{\lambda^2}{E_\pi}  \
H_{22}^{(3)}  \frac{\lambda^2}{E_\pi}  H_{21}^{(1)}  \
\frac{\lambda^1}{E_\pi}  H_{21}^{(1)}  \nn
&& {} -   H_{21}^{(1)}  \
\frac{\lambda^1}{E_\pi}  H_{21}^{(1)}  \frac{\lambda^2}{E_\pi^2}  \
H_{22}^{(3)}  \eta  H_{21}^{(1)}  \frac{\lambda^1}{E_\pi}  \
H_{21}^{(1)}   -   H_{21}^{(1)}  \frac{\lambda^1}{E_\pi}  \
H_{22}^{(3)}  \frac{\lambda^1}{E_\pi}  H_{21}^{(1)}  \eta  \
H_{21}^{(1)}  \frac{\lambda^1}{E_\pi^2}  H_{21}^{(1)}   \nn
&& {} + 2   \,
H_{21}^{(1)}  \frac{\lambda^1}{E_\pi}  H_{22}^{(3)}  \
\frac{\lambda^1}{E_\pi}  H_{21}^{(1)}  \frac{\lambda^2}{E_\pi}  \
H_{21}^{(1)}  \frac{\lambda^1}{E_\pi}  H_{21}^{(1)}    -   \
H_{21}^{(1)}  \frac{\lambda^1}{E_\pi}  H_{22}^{(3)}  \
\frac{\lambda^1}{E_\pi^2}  H_{21}^{(1)}  \eta  H_{21}^{(1)}  \
\frac{\lambda^1}{E_\pi}  H_{21}^{(1)}  \nn
&& {}  + 2  \, H_{21}^{(1)}  \
\frac{\lambda^1}{E_\pi}  H_{22}^{(3)}  \frac{\lambda^3}{E_\pi}  \
H_{21}^{(1)}  \frac{\lambda^2}{E_\pi}  H_{21}^{(1)}  \
\frac{\lambda^1}{E_\pi}  H_{21}^{(1)}   -   H_{21}^{(1)}  \
\frac{\lambda^1}{E_\pi^2}  H_{21}^{(1)}  \eta  H_{21}^{(1)}  \
\frac{\lambda^1}{E_\pi}  H_{21}^{(1)}  \frac{\lambda^2}{E_\pi}  \
H_{22}^{(3)}  \nn
&& {}  -  H_{21}^{(1)}  \frac{\lambda^1}{E_\pi^2}  \
H_{21}^{(1)}  \eta  H_{21}^{(1)}  \frac{\lambda^1}{E_\pi}  \
H_{22}^{(3)}  \frac{\lambda^1}{E_\pi}  H_{21}^{(1)}   -   \
H_{21}^{(1)}  \frac{\lambda^1}{E_\pi^2}  H_{21}^{(1)}  \eta  \
H_{22}^{(3)}  \frac{\lambda^2}{E_\pi}  H_{21}^{(1)}  \
\frac{\lambda^1}{E_\pi}  H_{21}^{(1)}  \nn
&& {} -   H_{21}^{(1)}  \
\frac{\lambda^1}{E_\pi^2}  H_{21}^{(1)}  \frac{\lambda^2}{E_\pi}  \
H_{22}^{(3)}  \eta  H_{21}^{(1)}  \frac{\lambda^1}{E_\pi}  \
H_{21}^{(1)}   -   H_{21}^{(1)}  \frac{\lambda^1}{E_\pi^2}  \
H_{22}^{(3)}  \frac{\lambda^1}{E_\pi}  H_{21}^{(1)}  \eta  \
H_{21}^{(1)}  \frac{\lambda^1}{E_\pi}  H_{21}^{(1)}  \nn
&& {}  -  \
H_{22}^{(3)}  \frac{\lambda^2}{E_\pi}  H_{21}^{(1)}  \
\frac{\lambda^1}{E_\pi}  H_{21}^{(1)}  \eta  H_{21}^{(1)}  \
\frac{\lambda^1}{E_\pi^2}  H_{21}^{(1)}   + 2\,  H_{22}^{(3)}  \
\frac{\lambda^2}{E_\pi}  H_{21}^{(1)}  \frac{\lambda^1}{E_\pi}  \
H_{21}^{(1)}  \frac{\lambda^2}{E_\pi}  H_{21}^{(1)}  \
\frac{\lambda^1}{E_\pi}  H_{21}^{(1)}  \nn
&& {}  -  H_{22}^{(3)}  \
\frac{\lambda^2}{E_\pi}  H_{21}^{(1)}  \frac{\lambda^1}{E_\pi^2}  \
H_{21}^{(1)}  \eta  H_{21}^{(1)}  \frac{\lambda^1}{E_\pi}  \
H_{21}^{(1)}   + 2 \,  H_{22}^{(3)}  \frac{\lambda^2}{E_\pi}  \
H_{21}^{(1)}  \frac{\lambda^3}{E_\pi}  H_{21}^{(1)}  \
\frac{\lambda^2}{E_\pi}  H_{21}^{(1)}  \frac{\lambda^1}{E_\pi}  \
H_{21}^{(1)}  \nn
&& {} -  H_{22}^{(3)}  \frac{\lambda^2}{E_\pi^2}  \
H_{21}^{(1)}  \frac{\lambda^1}{E_\pi}  H_{21}^{(1)}  \eta  \
H_{21}^{(1)}  \frac{\lambda^1}{E_\pi}  H_{21}^{(1)}  \bigg) \bigg] \eta \,, 
\eeqa
\item{terms $\propto g_A^2 c_i/m$:}
\beqa
V&=& \eta \bigg[
\alpha_9 \bigg(-  H_{20}^{(2)}  \eta  H_{21}^{(1)}  
\frac{\lambda^1}{E_\pi}  H_{21}^{(1)}  \frac{\lambda^2}{E_\pi^2}  
H_{22}^{(3)}    + H_{20}^{(2)}  \eta  H_{22}^{(3)}  
\frac{\lambda^2}{E_\pi^2}  H_{21}^{(1)}  \frac{\lambda^1}{E_\pi}  
H_{21}^{(1)}  +   H_{21}^{(1)}  \frac{\lambda^1}{E_\pi}  
H_{21}^{(1)}  \frac{\lambda^2}{E_\pi^2}  H_{22}^{(3)}  \eta  
H_{20}^{(2)} \nn
&&{}  -  H_{22}^{(3)}  \frac{\lambda^2}{E_\pi^2}  
H_{21}^{(1)}  \frac{\lambda^1}{E_\pi}  H_{21}^{(1)}  \eta  
H_{20}^{(2)}  \bigg) +\alpha_{10} \bigg(-  H_{20}^{(2)}  \eta  H_{21}^{(1)}
\frac{\lambda^1}{E_\pi^2}   H_{21}^{(1)}  \frac{\lambda^2}{E_\pi}  H_{22}^{(3)}   + 
  H_{20}^{(2)}  \eta  H_{22}^{(3)}  \frac{\lambda^2}{E_\pi}  
H_{21}^{(1)}  \frac{\lambda^1}{E_\pi^2}  H_{21}^{(1)}  \nn
&& {} +  
H_{21}^{(1)}  \frac{\lambda^1}{E_\pi^2}  H_{21}^{(1)}  
\frac{\lambda^2}{E_\pi}  H_{22}^{(3)}  \eta  H_{20}^{(2)}   - 
 H_{22}^{(3)}  \frac{\lambda^2}{E_\pi}  H_{21}^{(1)}  
\frac{\lambda^1}{E_\pi^2}  H_{21}^{(1)}  \eta  H_{20}^{(2)}  \bigg) +  
\alpha_{11} \bigg(-  H_{20}^{(2)}  \eta  H_{21}^{(1)}
\frac{\lambda^1}{E_\pi}  H_{22}^{(3)}  \frac{\lambda^1}{E_\pi^2}
H_{21}^{(1)}  \nn
&&{}  + 
  H_{20}^{(2)}  \eta  H_{21}^{(1)}  \frac{\lambda^1}{E_\pi^2}  
H_{22}^{(3)}  \frac{\lambda^1}{E_\pi}  H_{21}^{(1)}   +   
H_{21}^{(1)}  \frac{\lambda^1}{E_\pi}  H_{22}^{(3)}  
\frac{\lambda^1}{E_\pi^2}  H_{21}^{(1)}  \eta  H_{20}^{(2)}    - 
H_{21}^{(1)}  \frac{\lambda^1}{E_\pi^2}  H_{22}^{(3)}  
\frac{\lambda^1}{E_\pi}  H_{21}^{(1)}  \eta  H_{20}^{(2)}  \bigg) \nn
&& {} + \frac{1}{2} \bigg(  H_{20}^{(2)}  \eta  H_{21}^{(1)}  
\frac{\lambda^1}{E_\pi}  H_{21}^{(1)}  \frac{\lambda^2}{E_\pi^2}  
H_{22}^{(3)}   +   H_{20}^{(2)}  \eta  H_{21}^{(1)}  
\frac{\lambda^1}{E_\pi}  H_{22}^{(3)}  \frac{\lambda^1}{E_\pi^2}  
H_{21}^{(1)}   +   H_{20}^{(2)}  \eta  H_{21}^{(1)}  
\frac{\lambda^1}{E_\pi^2}  H_{21}^{(1)}  \frac{\lambda^2}{E_\pi}  
H_{22}^{(3)}  \nn
&& {} +   H_{20}^{(2)}  \eta  H_{21}^{(1)}  
\frac{\lambda^1}{E_\pi^2}  H_{22}^{(3)}  \frac{\lambda^1}{E_\pi}  
H_{21}^{(1)}   +   H_{20}^{(2)}  \eta  H_{22}^{(3)}  
\frac{\lambda^2}{E_\pi}  H_{21}^{(1)}  \frac{\lambda^1}{E_\pi^2}  
H_{21}^{(1)}   +   H_{20}^{(2)}  \eta  H_{22}^{(3)}  
\frac{\lambda^2}{E_\pi^2}  H_{21}^{(1)}  \frac{\lambda^1}{E_\pi}  
H_{21}^{(1)}  \nn
&& {}  - 2   H_{21}^{(1)}  \frac{\lambda^1}{E_\pi} 
H_{20}^{(2)}  \frac{\lambda^1}{E_\pi}  H_{21}^{(1)}  
\frac{\lambda^2}{E_\pi}  H_{22}^{(3)}   - 2  H_{21}^{(1)}  
\frac{\lambda^1}{E_\pi}  H_{20}^{(2)}  \frac{\lambda^1}{E_\pi}  
H_{22}^{(3)}  \frac{\lambda^1}{E_\pi}  H_{21}^{(1)}   - 2   
H_{21}^{(1)}  \frac{\lambda^1}{E_\pi}  H_{21}^{(1)}  
\frac{\lambda^2}{E_\pi}  H_{20}^{(2)}  \frac{\lambda^2}{E_\pi}  
H_{22}^{(3)}  \nn
&& {}  +   H_{21}^{(1)}  \frac{\lambda^1}{E_\pi}  
H_{21}^{(1)}  \frac{\lambda^2}{E_\pi^2}  H_{22}^{(3)}  \eta  
H_{20}^{(2)}   - 2   H_{21}^{(1)}  \frac{\lambda^1}{E_\pi}  
H_{22}^{(3)}  \frac{\lambda^1}{E_\pi}  H_{20}^{(2)}  
\frac{\lambda^1}{E_\pi}  H_{21}^{(1)}   +   H_{21}^{(1)}  
\frac{\lambda^1}{E_\pi}  H_{22}^{(3)}  \frac{\lambda^1}{E_\pi^2}  
H_{21}^{(1)}  \eta  H_{20}^{(2)}  \nn
&& {}  +   H_{21}^{(1)}  
\frac{\lambda^1}{E_\pi^2}  H_{21}^{(1)}  \frac{\lambda^2}{E_\pi}  
H_{22}^{(3)}  \eta  H_{20}^{(2)}   +   H_{21}^{(1)}  
\frac{\lambda^1}{E_\pi^2}  H_{22}^{(3)}  \frac{\lambda^1}{E_\pi}  
H_{21}^{(1)}  \eta  H_{20}^{(2)}   - 2   H_{22}^{(3)}  \frac{
\lambda^2}{E_\pi}  H_{20}^{(2)}  \frac{\lambda^2}{E_\pi}  
H_{21}^{(1)}  \frac{\lambda^1}{E_\pi}  H_{21}^{(1)}  \nn
&& {} - 2  
H_{22}^{(3)}  \frac{\lambda^2}{E_\pi}  H_{21}^{(1)}  
\frac{\lambda^1}{E_\pi}  H_{20}^{(2)}  \frac{\lambda^1}{E_\pi}  
H_{21}^{(1)}   +   H_{22}^{(3)}  \frac{\lambda^2}{E_\pi} 
H_{21}^{(1)}  \frac{\lambda^1}{E_\pi^2}  H_{21}^{(1)}  \eta  
H_{20}^{(2)}   +   H_{22}^{(3)}  \frac{\lambda^2}{E_\pi^2}  
H_{21}^{(1)}  \frac{\lambda^1}{E_\pi}  H_{21}^{(1)}  \eta  
H_{20}^{(2)}  \bigg) \bigg] \eta \,.
\eeqa
\end{itemize}
Here and in what follows, we adopt the notation of
Refs.~\cite{Epelbaum:2007us,Bernard:2007sp, Kolling:2009iq,Kolling:2011mt}. In particular, the subscripts $a$ and $b$ in
$H_{ab}^{(\kappa)}$  refer to the
number of the nucleon and pion fields, respectively, while the
superscript $\kappa$ gives the inverse mass dimension of the
corresponding coupling constant\footnote{For $1/m$-corrections,
  $\kappa_i$ corresponds to the inverse power of coupling constants
  plus twice the power of $m^{-1}$.  In particular, $\kappa=2$ for the
nucleon kinetic energy term $H_{20}$.},
 see Ref.~\cite{Epelbaum:2007us} for more details. 
The chiral
order associated with a given contribution can easily be read off by adding together the dimensions
$\kappa$  of $H_{ab}^{(\kappa)}$. More precisely, it is given by $\sum_i \kappa_i -2$.  
In the above equations,  $\eta$ ($\lambda$)  denote 
projection operators onto the purely nucleonic (the remaining) part of the
Fock space satisfying $\eta^2 = \eta$, $\lambda^2 = \lambda$, $ \eta \lambda 
= \lambda \eta = 0$ and $\lambda + \eta = {\bf 1}$. The superscript
$i$ of $\lambda^i$ refers to  the number of pions in the
corresponding intermediate state. Further, $E_\pi$ denotes the total
energy of the 
pions in the corresponding state, $E_\pi = \sum_i \sqrt{\vec l_i\, ^2 +
  M_\pi^2}$, with $\vec l_i$ the corresponding pion momenta.   Last but not least, 
the parameters $\alpha_{9, 10, 11}$ parametrize the unitary ambiguity of 
the resulting nuclear Hamiltonian.  

\def\theequation{\Alph{section}.\arabic{equation}}
\setcounter{equation}{0}
\section{Chiral expansion of the invariant $\pi N$ amplitudes $g^\pm (\omega, t)$ and $h^\pm (\omega, t)$}
\label{app2}

In this Appendix we give the explicit expressions for the invariant 
amplitudes $g^\pm (\omega, t)$ and $h^\pm (\omega, t)$ which parametrize 
the pion-nucleon scattering matrix at first four orders in the chiral expansion. 
We use here the heavy-baryon approach with the nucleon mass being counted 
according to $Q/m \sim Q^2 / \Lambda^2_\chi$. This has the consequence that 
relativistic corrections are shifted to higher orders as compared to the 
standard approach based on the assignment $m \sim \Lambda_\chi$. Apart 
from this difference, our results agree with the expressions given in Ref.~\cite{Fettes:2000xg}
(modulo one obvious misprint in that work).   

\underline{Contributions at order $Q$:}
\beq
g^{+}  =  0\,,\quad \quad
g^{-}  =  \frac{g_{A}^{2}\left(2M_{\pi}^{2}-t-2\,\omega^{2}\right)+2\,\omega^{2}}{4F_{\pi}^{2}\,\omega}\,,\quad \quad
h^{+}  =  -\frac{g_{A}^{2}}{2F_{\pi}^{2}\,\omega}\,,\quad \quad
h^{-}  =  0\,,
\eeq

\underline{Contributions at order $Q^2$:}
\beq
g^{+}  =  -\frac{4c_{1}M_{\pi}^{2} -2c_{2}\,\omega^{2} -c_{3}\left(2M_{\pi}^{2}-t\right)}{F_{\pi}^{2}}\,,
\quad \quad
g^{-}  =  0\,,\quad \quad
h^{+}  =  0\,,\quad \quad
h^{-}  =  \frac{c_{4}}{F_{\pi}^{2}}\,,
\eeq

\underline{Contributions at order $Q^3$:}
\begin{eqnarray}
g^{+} & = & \frac{i\sqrt{\,\omega^{2}-M_{\pi}^{2}}\left(g_{A}^{4}\left(M_{\pi}^{2}-\,\omega^{2}\right)\left(2M_{\pi}^{2}-t-2\,\omega^{2}\right)+3\,\omega^{4}\right)}{24\pi F_{\pi}^{4}\,\omega^{2}}-\frac{g_{A}^{2}\tilde{K}_{0}(t)\left(2M_{\pi}^{4}-5M_{\pi}^{2}t+2t^{2}\right)}{8F_{\pi}^{4}}\nonumber \\
 & + & \frac{g_{A}^{2}M_{\pi}\left(4g_{A}^{2}M_{\pi}^{2}\left(2M_{\pi}^{2}-t-2\,\omega^{2}\right)+3\,\omega^{2}\left(M_{\pi}^{2}-2t\right)\right)}{96\pi F_{\pi}^{4}\,\omega^{2}}-\frac{g_{A}^{2}\left(4M_{\pi}^{4}-4M_{\pi}^{2}t+t\left(t+4\,\omega^{2}\right)\right)}{16F_{\pi}^{2}m\,\omega^{2}}\,,\nonumber \\[3pt]
g^{-} & = & \frac{(\bar{d}_{1}+\bar{d}_{2})\left(4M_{\pi}^{2}\,\omega-2t\,\omega\right)}{F_{\pi}^{2}}+\frac{4\bar{d}_{3}\,\omega^{3}}{F_{\pi}^{2}}+\frac{8\bar{d}_{5}M_{\pi}^{2}\,\omega}{F_{\pi}^{2}}+\frac{\bar{d}_{18}g_{A}M_{\pi}^{2}\left(-2M_{\pi}^{2}+t+2\,\omega^{2}\right)}{F_{\pi}^{2}\,\omega}\nonumber \\
 & + & \frac{\tilde{J}_{0}(\omega)\left(g_{A}^{4}\left(M_{\pi}^{2}-\,\omega^{2}\right)\left(2M_{\pi}^{2}-t-2\,\omega^{2}\right)+6\,\omega^{4}\right)}{12F_{\pi}^{4}\,\omega^{2}}+\frac{i\sqrt{\,\omega^{2}-M_{\pi}^{2}}\left(g_{A}^{4}\left(M_{\pi}^{2}-\,\omega^{2}\right)\left(2M_{\pi}^{2}-t-2\,\omega^{2}\right)+6\,\omega^{4}\right)}{96\pi F_{\pi}^{4}\,\omega^{2}}\nonumber \\
 & + & \frac{g_{A}^{4}\left(3M_{\pi}^{2}-2\,\omega^{2}\right)\left(2M_{\pi}^{2}-t-2\,\omega^{2}\right)-g_{A}^{2}\,\omega^{2}\left(12M_{\pi}^{2}+t\right)+\,\omega^{2}\left(t-6M_{\pi}^{2}\right)}{288\pi^{2}F_{\pi}^{4}\,\omega}\nonumber \\
 & + &\frac{\tilde{I}_{20}(t)\,\omega\left(-4\left(2g_{A}^{2}+1\right)M_{\pi}^{2}+5g_{A}^{2}t+t\right)}{12F_{\pi}^{4}} -\frac{\left(-4M_{\pi}^{2}+t+4\,\omega^{2}\right)\left(g_{A}^{2}\left(2M_{\pi}^{2}-t+2\,\omega^{2}\right)-2\,\omega^{2}\right)}{16F_{\pi}^{2}m\,\omega^{2}}\,,\nonumber \\[3pt]
h^{+} & = & \frac{2(\bar{d}_{14}-\bar{d}_{15})\,\omega}{F_{\pi}^{2}}+\frac{2\bar{d}_{18}g_{A}M_{\pi}^{2}}{F_{\pi}^{2}\,\omega}+\frac{g_{A}^{4}\tilde{J}_{0}(\omega)\left(\,\omega^{2}-M_{\pi}^{2}\right)}{6F_{\pi}^{4}\,\omega^{2}}+\frac{ig_{A}^{4}\left(\,\omega^{2}-M_{\pi}^{2}\right)^{3/2}}{48\pi F_{\pi}^{4}\,\omega^{2}}-\frac{g_{A}^{4}\left(3M_{\pi}^{2}+4\,\omega^{2}\right)}{144\pi^{2}F_{\pi}^{4}\,\omega}+\frac{g_{A}^{2}\left(t-4M_{\pi}^{2}\right)}{8F_{\pi}^{2}m\,\omega^{2}}\,,\nonumber \\[3pt]
h^{-} & = & \frac{ig_{A}^{4}\left(\,\omega^{2}-M_{\pi}^{2}\right)^{3/2}}{24\pi F_{\pi}^{4}\,\omega^{2}}+\frac{g_{A}^{2}\tilde{K}_{0}(t)\left(4M_{\pi}^{2}-t\right)}{8F_{\pi}^{4}}-\frac{4g_{A}^{4}M_{\pi}^{3}+3g_{A}^{2}M_{\pi}\,\omega^{2}}{96\pi F_{\pi}^{4}\,\omega^{2}}+\frac{g_{A}^{2}\left(2M_{\pi}^{2}-t-2\,\omega^{2}\right)+2\,\omega^{2}}{8F_{\pi}^{2}m\,\omega^{2}}\,,
\end{eqnarray}

\underline{Contributions at order $Q^4$:}
\begin{eqnarray}
g^{+} & = & \frac{2c_{1}\tilde{I}_{20}(t)M_{\pi}^{2}\left(M_{\pi}^{2}-2t\right)}{F_{\pi}^{4}}+c_{2}\left(-\frac{\tilde{I}_{20}(t)\left(4M_{\pi}^{4}-9M_{\pi}^{2}t+ 2t^{2}\right)}{12F_{\pi}^{4}}-\frac{6M_{\pi}^{4}-13M_{\pi}^{2}t+2t^{2}}{288\pi^{2}F_{\pi}^{4}}+ \frac{\,\omega\left(-4M_{\pi}^{2}+t+4\,\omega^{2}\right)}{F_{\pi}^{2}m}\right)\nonumber \\
 & - & \frac{c_{3}\tilde{I}_{20}(t)\left(2M_{\pi}^{4}-5M_{\pi}^{2}t+2t^{2}\right)}{2F_{\pi}^{4}}+\frac{4\bar{e}_{14}\left(t- 2M_{\pi}^{2}\right)^{2}}{F_{\pi}^{2}}+\frac{8\bar{e}_{15}\,\omega^{2}\left(2M_{\pi}^{2}-t\right)}{F_{\pi}^{2}}+ \frac{16\bar{e}_{16}\,\omega^{4}}{F_{\pi}^{2}}\nonumber \\ 
 & + & \frac{4M_{\pi}^{2}\left(2\bar{e}_{19}-\bar{e}_{22}-\bar{e}_{36}+2 \bar{l}_{3}c_{1} F_{\pi}^{-2} \right)\left(2M_{\pi}^{2}-t\right)}{F_{\pi}^{2}}+\frac{16\left(\bar{e}_{20}+\bar{e}_{35}\right)M_{\pi}^{2}\, \omega^{2}}{F_{\pi}^{2}}+\frac{8\left(\bar{e}_{22}-4\bar{e}_{38}-\bar{l}_{3}c_{1}  F_{\pi}^{-2}\right)M_{\pi}^{4}}{F_{\pi}^{2}}\,,\nonumber \\[3pt]
g^{-} & = & -\frac{ic_{1}M_{\pi}^{2}\,\omega\sqrt{\,\omega^{2}-M_{\pi}^{2}}}{\pi F_{\pi}^{4}}+\frac{ic_{2}\,\omega^{3}\sqrt{\,\omega^{2}-M_{\pi}^{2}}}{2\pi F_{\pi}^{4}}\nonumber \\
 & + & c_{3}\left(\frac{g_{A}^{2}M_{\pi}^{3}\left(2M_{\pi}^{2}-t-2\,\omega^{2}\right)}{12\pi F_{\pi}^{4}\,\omega}+\frac{i\sqrt{\,\omega^{2}-M_{\pi}^{2}}\left(g_{A}^{2}\left(M_{\pi}^{2}-\,\omega^{2}\right) \left(2M_{\pi}^{2}-t-2\,\omega^{2}\right)+6\,\omega^{4}\right)}{12\pi F_{\pi}^{4}\,\omega}\right)\nonumber \\
 & + & c_{4}\left(-\frac{ig_{A}^{2}\left(\,\omega^{2}-M_{\pi}^{2}\right)^{3/2}\left(-2M_{\pi}^{2}+t+2\, \omega^{2}\right)}{12\pi  F_{\pi}^{4}\,\omega}+\frac{g_{A}^{2}M_{\pi}^{3}\left(-2M_{\pi}^{2}+t+2\,\omega^{2}\right)}{12\pi F_{\pi}^{4}\,\omega}+\frac{t\,\omega}{2F_{\pi}^{2}m}\right)\,,\nonumber \\[3pt]
h^{+} & =( & c_{3}-c_{4})\left(\frac{ig_{A}^{2}\left(\,\omega^{2}-M_{\pi}^{2}\right)^{3/2}}{6\pi F_{\pi}^{4}\,\omega}-\frac{g_{A}^{2}M_{\pi}^{3}}{6\pi F_{\pi}^{4}\,\omega}\right)\,,\nonumber \\[3pt]
h^{-} & = & c_{4}\Bigg(\frac{4g_{A}^{2}\tilde{J}_{0}(\omega)\left(M_{\pi}^{2}-\,\omega^{2}\right)}{3F_{\pi}^{4}\, \omega}+\frac{-6\left(5g_{A}^{2}+1\right)M_{\pi}^{2}+8g_{A}^{2}\,\omega^{2}+t}{144\pi^{2}F_{\pi}^{4}}+ \frac{ig_{A}^{2}\left(M_{\pi}^{2}-\,\omega^{2}\right)\sqrt{\,\omega^{2}-M_{\pi}^{2}}}{6\pi F_{\pi}^{4}\,\omega}\nonumber \\
 & + & \frac{\tilde{I}_{20}(t)\left(t-4M_{\pi}^{2}\right)}{6F_{\pi}^{4}}+\frac{\,\omega}{F_{\pi}^{2}m}\Bigg)+ \frac{\bar{e}_{17}\left(8M_{\pi}^{2}-4t\right)}{F_{\pi}^{2}}+\frac{8\bar{e}_{18}\,\omega^{2}}{F_{\pi}^{2}}+ \frac{8\left(\bar{e}_{21}-\frac{\bar{e}_{37}}{2}\right)M_{\pi}^{2}}{F_{\pi}^{2}}\,. 
\label{piNQ4}
\end{eqnarray}
The low-energy constants and the kinematical variables entering these expressions 
are defined in sections \ref{sec:lagr}, \ref{sec:TPE} and \ref{sec:piN}. The loop functions 
are defined via:
\begin{eqnarray}
\tilde{J}_{0}(\omega) & = &\frac{\omega}{8\pi^2} -\frac{\sqrt{\,\omega^{2}-M_{\pi}^{2}}}{4\pi^{2}}
\log\left(\sqrt{\frac{\,\omega^{2}}{M_{\pi}^{2}}-1}+ \frac{\,\omega}{M_{\pi}}\right)\,, \nn [3pt]
\tilde{K}_{0}(t)&=&-\frac{1}{8\pi\sqrt{-t}} \arctan \frac{\sqrt{-t}}{2M_{\pi}}\,,\nn [3pt]
\tilde{I}_{20}(t) & = & \frac{1 }{16\pi^{2}} - \frac{\sqrt{1-4M_{\pi}^{2}/t}}{16\pi^{2}}
\log\frac{\sqrt{1-4M_{\pi}^{2}/t}+1}{\sqrt{1- 4M_{\pi}^{2}/t}-1}\,.
\end{eqnarray}

\end{document}